\documentclass[aps,prb,twocolumn,superscriptaddress,showpacs,floatfix]{revtex4}

\usepackage[bookmarks]{hyperref}
\usepackage[dvips]{graphicx}
\usepackage{amsmath}
\usepackage{amstext}
\usepackage{graphics}
\usepackage{exscale}
\usepackage{epsfig}
\usepackage[T1]{fontenc}
\usepackage{bm}
\usepackage{bbm}
\usepackage{color}

\usepackage{version}

\usepackage{latexsym}

\usepackage[tight]{subfigure}

\renewcommand{\epsilon}{\varepsilon}

\newcommand{\integral}[3]{\!\int\limits_{#2}^{#3}\!\!{\rm d}#1\;}

\newcommand{\elcre}[2]{ c^{\dagger}_{#1,#2}}
\newcommand{\elann}[2]{ c_{#1,#2}}

\newcommand{\e}{\mathrm e}

\newcommand{\vct}[1]{\bm #1}
\newcommand{\vk}{{\bm k}}
\newcommand{\vq}{{\bm q}}

\newcommand{\vkF}{\vk_{{\scriptscriptstyle \mathrm{F}}}}

\newcommand{\Imag}{\mathrm{Im}}
\newcommand{\Real}{\mathrm{Re}}

\newcommand{\hc}{\mathrm{h.c.}}

\includeversion{commentst1} 

\begin{document}

\title{Local origin of the pseudogap in the attractive Hubbard model} 
\author{Robert Peters}
\email[]{robert.peters@riken.jp}
\affiliation{Computational Condensed Matter Physics Laboratory, RIKEN, Wako, Saitama 351-0198, Japan}
\author{Johannes Bauer}
\email[]{jbauer@physics.harvard.edu}
\affiliation{Department of Physics, Harvard University, Cambridge,
  Massachusetts 02138, USA}
\date{\today} 

\begin{abstract}
We provide a new perspective on the pseudogap physics for attractive fermions
as described by the three-dimensional Hubbard model. The pseudogap in the
single-particle spectral function, which occurs for temperatures above the critical
temperature $T_c$ of the superfluid transition, is often interpreted in terms of preformed,
uncondensed pairs. Here we show that the occurrence of pseudogap physics can be consistently
understood in terms of local excitations which lead to a splitting of the
quasiparticle peak for sufficiently large interaction. This effect becomes prominent
at intermediate and high temperatures when the quantum mechanical hopping is
incoherent. We clarify the existence of a conjectured temperature below which
pseudogap physics is expected to occur. Our results are based on 
approximating the physics of the three-dimensional Hubbard model by dynamical
mean field theory calculations and a momentum independent self-energy.
Our predictions can be tested with ultracold atoms in optical lattices with
currently available temperatures and spectroscopic techniques.
\end{abstract}
\pacs{74.20.-z,74.25.Dw,74.25.Gz,71.10.Fd,67.85.-d,67.25.dj}

%

\maketitle

\section{Introduction} 

The analysis of peaks in the single particle spectral function, measured, for
instance, by photoemission experiments in solids or radio frequency (RF)
spectroscopy for ultracold atoms, provides important information about
correlation effects in interacting quantum many-body systems. 
In the limit of weak interactions the spectral function displays peaks close
to the energies of the free fermion energy-momentum distribution and as such
directly represents single-particle properties. 
At finite temperature and energies away from the Fermi surface, peaks are
broadened and the width is indicative of interaction effects, which open up
decay channels. If spontaneous symmetry breaking occurs below a certain 
temperature, such as in a superconductor below $T_c$, the single-particle
excitations become gapped out and shifted by an amount $\Delta$, the
superconducting gap. 

A more peculiar behavior is that of excitations being gapped out
(or suppressed) even though no obvious symmetry breaking and thermodynamic ordering
transition occurs. This is often referred to as pseudogap (PG) physics. The
spectral gap can look very similar to a gap due to symmetry breaking at 
finite temperature. Therefore, it can be difficult to clarify the origin of
PG physics and to distinguish whether it is due to some hidden order or
a different effect. A very prominent example of such physics is provided by
the experimental observations in the hole doped copper-oxide high temperature
superconductors,\cite{TS99c,Kor15} where a relatively 
large part of the phase diagram is occupied by such a PG
behavior. This phenomenon has attracted an enormous amount of attention,
however, there is currently no consensus about the physical origin of the
this PG for the cuprates, and different scenarios have been invoked as
an explanation. These include hidden order,\cite{CLMN01} spin-fluctuations, \cite{Sca12}
phase fluctuations and preformed pairs,\cite{EK95,MCCN14}
and the interplay with charge fluctuations.\cite{EMP13}

Here we focus on a conceptually simpler situation where PG physics has
also been reported and that is for systems of fermions with locally attractive 
interactions. In situations without nesting the dominant instability at low
temperature is superconductivity and, correspondingly, pairing processes are
expected to be most relevant. In particular, the crossover from weak coupling
\citet*{BCS57} (BCS) theory  
of superconductivity to strong coupling Bose-Einstein condensation (BEC) of pairs
has been studied extensively and is a classical problem in condensed
matter physics.\cite{Eag69,Leg80,NS85,Ran95} In the last decade it has
attracted renewed interest due to experimental realizations with ultracold fermions.
Superfluidity has been reported in such
systems,\cite{GRJ03,ZSSRKK04,ZSSSK05,BDZ08} also in the case where the
fermions are confined to an optical lattice.\cite{CMLSSSXK06} Moreover, based
on RF spectroscopy PG signatures have been reported for two \cite{FFVKK11} and
three-dimensional systems without optical lattices.\cite{GSDJPPS10,Ran10} 

\begin{figure}
\begin{center}
\includegraphics[width=0.95\columnwidth]{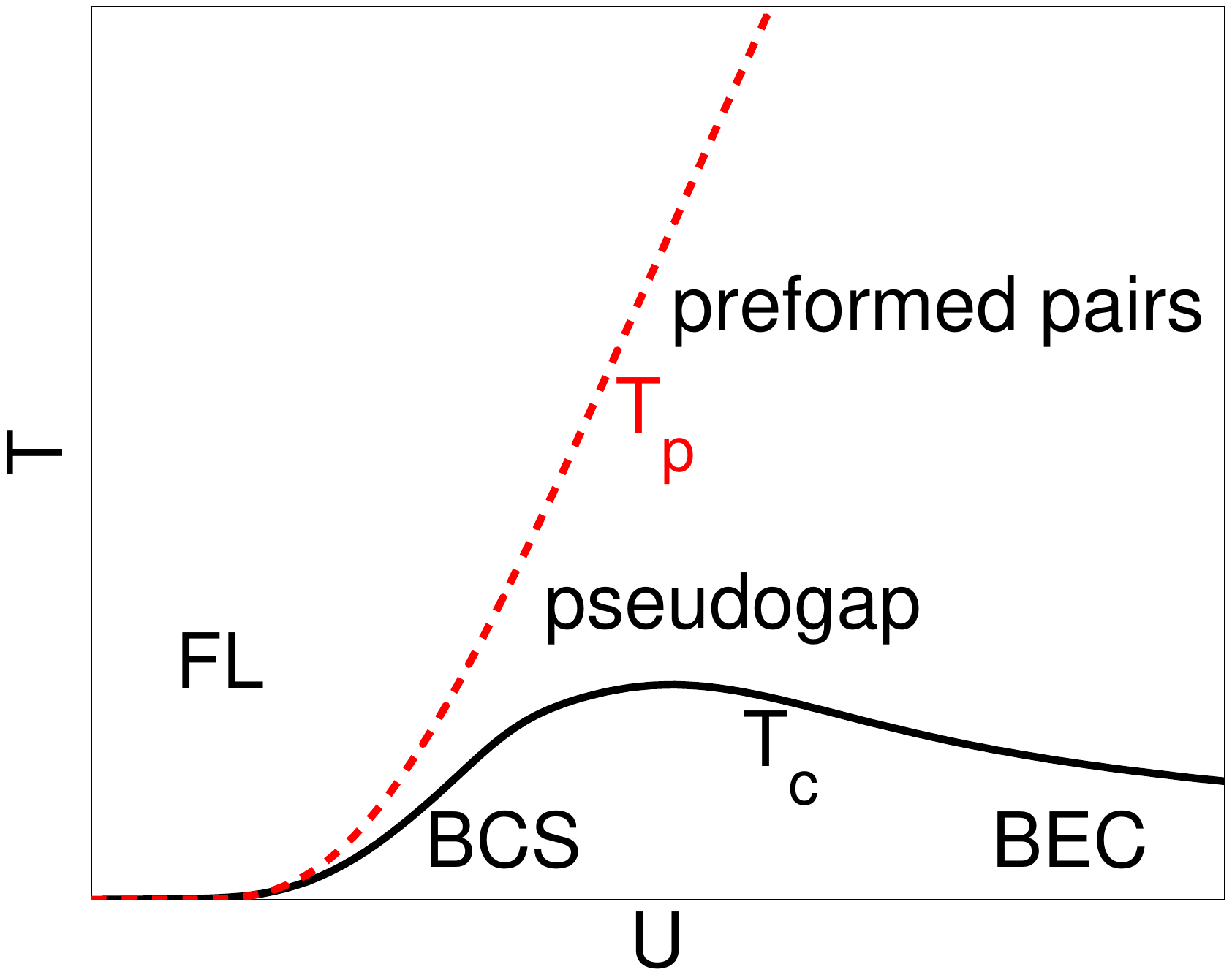}
\end{center}
\vspace{-0.5cm}
\caption{(Color online) Schematic phase diagram for the preformed pair scenario in the $T-U$
  plane with critical temperature $T_c$, pairing temperature $T_{\rm p}$,
  Fermi liquid (FL) regime, and PG physics below $T_{\rm p}$ (after
  Randeria \cite{Ran10}). 
\label{schemphase_diagram}}
\end{figure}

Three non-exclusive concepts are usually invoked to discuss the origin of 
PG physics for attractive fermions: (i) {\em Preformed pairs}; for
intermediate coupling strength, pair formation without condensation is expected
to occur at a certain temperature $T_{\rm p}$ which is larger than the
superfluid (SF) phase transition temperature $T_c$. These preformed pairs can lead to PG
formation as a certain binding energy is required to break the pair and resolve a single
fermion excitation.\cite{Ran95,Ran10,CLS06} This idea leads to a popular scenario for PG
physics and is illustrated in a schematic phase diagram in
Fig.~\ref{schemphase_diagram}. (ii) {\em Pairing fluctuations} above $T_c$ and
their 
effect on single particle properties via a many-body self-energy can lead to
PG physics.\cite{CLS06} 
(iii) {\em Phase fluctuations}; in a situation where fermions
are paired one can imagine that a finite magnitude of the order parameter 
establishes locally, however, no macroscopic coherent SF phase
develops due to strong phase fluctuation.\cite{EK95}  In this situation the presence of
the ordering tendency related to a gap can then lead to PG signatures in the
spectral functions.\cite{ESAH02} 
This behavior, which coincides with a small SF
stiffness, is expected to be particularly pronounced in two-dimensional systems.

Within one and the same calculation it is very difficult to obtain
non-perturbative results {\em and} to include all relevant fluctuation effects.
The purpose of this work is to contribute to a better understanding of the
importance of particular effects in the lattice situation based on
non-perturbative calculations. 
It is important to distinguish different setups when comparing the
occurrence of PG physics for attractive fermions. First of all, dimensionality
plays an important role in determining the strength of fluctuations, and in
particular, the two-dimensional situation has more pronounced fluctuation effects. 
Moreover, results can differ in calculations for a model defined in the continuum
and one on a lattice, such as the Hubbard model. 
A well known example is the $T_c$ curve which drops with the coupling strength on the
lattice as $1/U$, whereas it approaches a constant in the continuum. 
Here we will analyze the attractive Hubbard model in three spatial
dimensions. We will use the dynamical mean field theory (DMFT) approximation
\cite{GKKR96} to compute the self-energies and spectral function in the normal
and SF phase. This approximation is non-perturbative in the
interaction strength and therefore can describe very well the occurrence of
preformed pairs. However, it does not include the effect of phase fluctuations
(iii) and also does  not include the effect of small momentum pairing
fluctuations. The PG physics observed in our work can therefore not be related
to such effects. Phase fluctuations above $T_c$ are usually argued to be of
minor importance for spectral properties in three dimensions.

\begin{figure}
\begin{center}
\includegraphics[width=0.95\columnwidth]{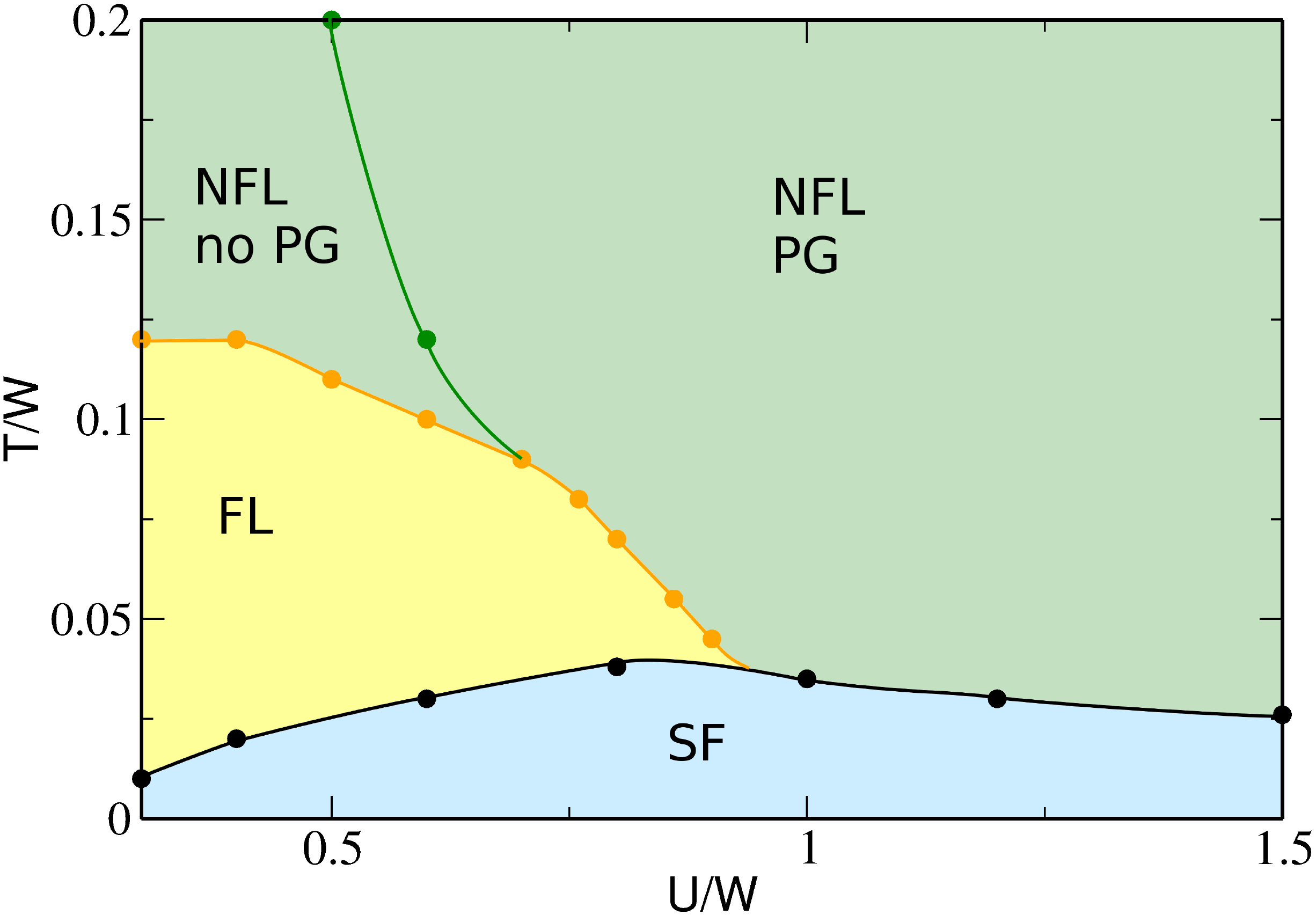}
\end{center}
\caption{(Color online) Phase diagram at half filling. We distinguish four
  different regimes: the superfluid phase (SF); a 
  non-Fermi-liquid regime (NFL), which is separated into a region with
  PG in $\rho(\omega)$ and a region without PG (no PG); and a Fermi liquid regime (FL) below the
  temperature $T_{\rm FL}$.
\label{phase_diagram}}
\end{figure}

There is a substantial literature of previous work on BCS-BEC crossover and PG
physics for attractive fermions, which however does not provide a clear and
complete picture about PG physics.
A popular approach is the diagrammatic T-matrix approximation,\cite{CSTL05,CLS06,CGHL10}
which captures well the effect of pairing fluctuations (ii). It was applied to the 2d
Hubbard model \cite{KMS99,RM01} and PG features have been found in the {\em
  non}-selfconsistent version,\cite{CLS06} also in the continuum in two
\cite{MPPPS15} and three dimensions.\cite{PPSC02,TWO09,WTO10} Selfconsistent
T-matrix calculations for the 
3d continuum model have found no PG in the spectrum.\cite{Hau92,HPZ09}
However, in the two-dimensional case recently PG behavior was found.\cite{BME14}
There are also non-perturbative calculations, such as DMFT and quantum Monte Carlo (QMC)
which found PG features in the continuum model\cite{MWBD09,HLDD10,WMDBR13} and for the Hubbard
model at different filling factors.\cite{RTMS92,TR95,VT97,MALKPVT00,KW11a,KW11b,RT14} The latter
results were found to be in good agreement with a diagrammatic technique.\cite{KAT01} It is worth
noting that QMC techniques usually need to perform
analytic continuation of imaginary axis data which can lead to uncertainties
in results for spectral functions.
DMFT studies, including cellular versions, for the
attractive Hubbard model have been carried out in the normal
phase,\cite{KMS01,CCG02,KGT06,KKS14} and in the broken symmetry
phase.\cite{GKR05,TBCC05,TCC05,BH09,BHD09,KW11a,KW11b}

Our major results are the following:
\begin{itemize}
\item 
 For large enough coupling strength we find PG physics at temperatures
 $T>T_c$. At half filling the PG remains for {\em all} temperatures above
 $T_c$ and therefore a pairing temperature $T_{\rm p}$
 (Fig.~\ref{schemphase_diagram}) is not decisive to invoke the PG
in the spectral function (see Fig.~\ref{phase_diagram}). For different fillings the spectral function is
shifted due to the flattening of the Fermi function, such that the main
suppression of spectral weight does not occur at $\omega=0$. 
\item 
 The occurrence of PG physics at high temperatures can be understood via split
 local excitations on lattice sites visible for strong enough interactions. 
\item
 PG physics in the spectral function is related to Non Fermi Liquid (NFL) properties
 of the self-energy (see Fig.~\ref{fig:sigma_schematic}, detailed definition below). 
\item
We demonstrate in detail how the PG transforms smoothly into the
superconducting gap, when the temperature is lowered through $T_c$ (see
Fig.~\ref{TC}).  
\end{itemize}
The paper is organized as follows: In Sec.~II we briefly describe our model
and method. Sec.~III discusses conceptual background about the occurrence of
PG physics in relation to the self-energy. In Sec.~IV and V we show results
for spectra and self-energies at and away from half filling before concluding
in Sec.~VI. In the appendix we compare the DMFT-NRG calculations to iterated
perturbation theory and to T-matrix calculations.

\section{Model definition and DMFT calculations}
Our study is based on the three-dimensional attractive Hubbard
model,\cite{MRR90,Ran95} which in the grand canonical formalism reads  
\begin{equation}
H=\sum_{i,j,\sigma}(t_{ij}\elcre {i}{\sigma}\elann
{j}{\sigma}+\hc)-\mu\sum_{i\sigma}n_{i\sigma}-U\sum_in_{i,\uparrow}n_{i,\downarrow},
\label{attHub}
\end{equation}
with the chemical potential $\mu$, the interaction strength $U>0$ and  the
hopping parameters $t_{ij}$. $\elcre {i}{\sigma}$ creates a fermion at site
$i$ with spin $\sigma$, and $n_{i,\sigma}=\elcre {i}{\sigma}\elann
{i}{\sigma}$.
We take only a nearest neighbor hopping ($-t$), so that the non-interacting
energy-dispersion relation in the three-dimensional cubic system is given as,
$\vk=(k_x,k_y,k_z)$, 
\begin{equation}
  \epsilon_{\vk}=-2t[\cos(k_x)+\cos(k_y)+\cos(k_z)].
\label{eq:dispersion}
\end{equation}
The dispersion satisfies $\epsilon_{-\vk}= \epsilon_{\vk}$. The corresponding
density of states (DOS) is denoted by $\rho_0(\epsilon)$. In the calculations we
will use the hopping $t$ and the bandwidth $W=12t$ as energy scales.

The main method used to study the Hamiltonian (\ref{attHub}) is the dynamical
mean field theory (DMFT).\cite{GKKR96}
Within DMFT, we have to self-consistently solve a quantum impurity model 
describing a single lattice site in the environment of all other lattice
sites. In order to calculate the self energy for this quantum impurity model,
we mainly use the numerical renormalization group (NRG),\cite{BCP08} which is able to
calculate accurately expectation values, Green's functions, and self energies at
zero and finite temperatures\cite{PPA06,WD07} also in the superconducting
case.\cite{BOH07,HWDB08,BH09,BHD09} 
Dynamical correlation functions are calculated within the NRG by broadening of
a large  number of discrete excitations in the Lehman representation, and as
such do not require analytic continuation. 
For our calculations, we choose a log-normal broadening function \cite{BCV01,BCP08} with
unusually narrow and temperature-independent width, $b=0.3$. One of the main reasons
for this is to avoid a large transfer of spectral weight to high energies
which can be particularly important at higher temperatures.
Using this narrow broadening leads to artificial oscillations in the spectra,
which originate from the discretization of the bath in the
NRG-calculation. In order to produce physical spectra, we finally smooth these
oscillations by averaging over $\Delta\omega=0.01 W$. This averaging is
justified for the present purpose, because  we do not expect very fine and sharp
structures in our spectra on this energy scales to determine the physics of
the PG. Furthermore, we carefully compared our NRG calculated spectra with
iterated perturbation theory (see appendix). The latter technique does not
require to broaden discrete excitations and provides therefore a useful test
in a suitable parameter regime.

\section{Features of the spectral functions and self-energy}
Before presenting the results of our calculations it is useful to discuss some
basic features of the Green's functions and self-energy, which will help
us to better understand under which conditions PG physics occurs.  In the literature
PG physics is considered quite generally either for the
integrated spectral function $\rho(\omega)=\frac{1}{N}\sum_{\vk}\rho_{\vk}(\omega)$,
which is equivalent to the local spectrum $\rho_{ii}(\omega)$, or for $\vk$-resolved spectra
$\rho_{\vk}(\omega)$ close to the Fermi surface. We will consider both
quantities in this paper. We note that a PG in one of
these does not necessarily imply one in the other quantity.

Let us first note that in the limit of high temperature $T\gg W,U$
correlation lengths become small and the physics is dominated by local
processes.\cite{Geo11} 
This is seen, for instance, when we consider the bare single-particle propagator in
imaginary time 
\begin{equation}
G_{ij}^0(\tau)=-\frac{1}{N}\sum_{\vk}\e^{i\vk \vct r_{ij}}\e^{-\xi_{\vk}\tau}
\e^{\beta\xi_{\vk}} n_{\rm F}(\xi_{\vk}),  
\label{eq:G0tau}
\end{equation}
where $\xi_{\vk}=\epsilon_{\vk}-\mu$ and $\tau\in[0,\beta)$. In the limit of
high temperature, $\beta=1/T\to 0$ and $n_{\rm F}(\xi_{\vk})\to 1/2$. Then
$G_{ij}^0(\tau)$ in Eq.~(\ref{eq:G0tau}) becomes essentially local, $\sim
\delta_{ij}$, and spatial components with $i\neq j$ vanish exponentially with
length scale $\lambda\sim \frac{at}{T}$.\cite{fn1} 
Quantum mechanical
hopping is largely incoherent in this situation. 
This also means that DMFT based on a local self-consistent approximation
can become very accurate in this high temperature limit.
For instance, high temperature expansions for the three dimensional Hubbard
model agree well with DMFT calculations for thermodynamical quantities, and
this remains to be the case down to temperatures of the order $T\simeq
W/8$.\cite{Geo11,LBKGS11} 
One should, however, note that the self-energy of the three dimensional
Hubbard model does not become completely $\vk$-independent even in the limit
$T\to\infty$.\cite{KPRS14}

What are the implications from this for the spectral function and the self-energy?
The excitations in the limit where local physics dominates
are determined by the local part of the Hamiltonian,  $H_{\rm
loc}=-\mu\sum_{i\sigma}n_{i\sigma}-U\sum_in_{i,\uparrow}n_{i,\downarrow}$. At
half filling the chemical potential is fixed to $\mu=-U/2$, and depending on
the occupation $n=0,1,2$ we have the energies $E_{\alpha}=0,U/2,0$,
respectively. Excitations in the spectral function have finite matrix
elements for states where the particle number differs by one. Hence, in the spectral
function excitation at energies $\Delta E=\pm U/2$ can be expected. The corresponding
self-energy for the atomic problem reads,
$\Sigma_{ii}(\omega)=\frac{U^2}{4(\omega+i\Gamma)}$, where $\Gamma\to 0$. This
implies $\delta$-function peaks at $\pm U/2$ in the spectral function. In
Sec.~IV we will see that DMFT results at high temperature and large $U$ are
indeed of a similar form,  
$\Imag\Sigma_{ii}(\omega)=-\frac{U^2\Gamma}{4(\omega^2+\Gamma^2)}$. 
Away from half filling the situation is more complicated, but similar
features remain visible.
If the peak in the self-energy is strong enough, 
we find in the spectral function increased weight at $\omega=\pm U/2$ and a suppression of
spectral weight at the Fermi energy. These are the signatures of the PG in the
integrated spectral function. 
For strong interactions this effect remains observable down to
intermediate temperatures. In other words, the PG in $\rho(\omega)$ is
related to the existence of Hubbard bands which are visible in the spectral
function at all temperatures.

We now discuss the appearance of a gap and PG in the momentum resolved
spectral function $\rho_{\vk}(\omega)$.
In the normal phase the Matsubara Green's function reads 
\begin{equation}
  G_{\vk}(i\omega_n)=\frac{1}{i\omega_n-\xi_{\vk}-\Sigma(i\omega_n)},
\end{equation}
where we have assumed a momentum
independent self-energy as appropriate for DMFT calculations. 
The spectral function is obtained from analytic continuation, $i\omega_n\to
\omega+i\eta$, $\eta\to 0$, to yield
\begin{equation}
  \rho_{\vk}(\omega)=-\frac{1}{\pi}\frac{\Sigma^I(\omega)}{[\omega-\xi_{\vk}-\Sigma^R(\omega)]^2+\Sigma^I(\omega)^2}. 
\end{equation}
We have separated real (R) and imaginary (I) parts of the self-energy.

In the SF state we can include an explicit symmetry breaking term,
$\Delta_{\rm  sc}^{0}$, $\Delta_{\rm  sc}^{0}\to 0$ for spontaneous symmetry
breaking, and 
the non-interacting Green's function matrix $\underline G_{\vk}^0(i\omega_n)$ has
the form,     
\begin{equation}
  \underline G_{\vk}^0(i\omega_n)^{-1}=
\left(\begin{array}{cc}
i\omega_n-\xi_{\vk}  & \Delta^0_{\rm sc}  \\
\Delta^0_{\rm sc} & i\omega_n + \xi_{\vk}
\end{array}\right),
\label{frGfctsc}
\end{equation}
For the interacting system we introduce the matrix self-energy $\underline\Sigma_{\vk}(i\omega_n)$ such
that the inverse of the full Green's function matrix $\underline
G_{\vk}(i\omega_n)$ is given by the Dyson equation  
\begin{equation}
  \underline G_{\vk}(i\omega_n)^{-1}=
  \underline G_{\vk}^0(i\omega_n)^{-1}-\underline\Sigma_{\vk}(i\omega_n).
\label{scdyson}
\end{equation}
The diagonal component of the $\vk$-dependent Green's function reads
\begin{equation}
G_{\vk}(\omega)=\frac{\zeta_{2,\vk}(i\omega_n)}{\zeta_{1,\vk}(i\omega_n)\zeta_{2,\vk}(i\omega_n)-\Sigma_{12}(i\omega_n)\Sigma_{21}(i\omega_n)} ,
\end{equation}
with $\zeta_{1,\vk}(i\omega_n)=\omega-\xi_{\vk}-\Sigma_{11}(i\omega_n)$,
$\zeta_{2,\vk}(i\omega_n)=i\omega_n+\xi_{\vk}-\Sigma_{22}(i\omega_n)$.
The off-diagonal self-energy $\Sigma_{12}(\omega)$, in particular its real
part, plays the role of a dynamic gap function,
$\Real\Sigma_{12}(\omega)\sim
\Delta$. Therefore, low energy spectral excitations which correspond to
$\omega=z(\xi_{\vk}-\Sigma^{R}(0))$ in the normal phase are  
shifted by the gap $\Delta$ to $\pm
E_{\vk}\sim\pm z\sqrt{(\xi_{\vk}-\Sigma^{R}(0))^2+\Delta^2}$, where
$z^{-1}=1-\partial_{\omega}\overline{\Sigma}^R_{11}(0)$ is the renormalization
factor. Usually we associate the gap with a
binding energy of pairs and hence we can interpret this energy shift as an energy
required to break a pair and see a single-particle excitation. 

\begin{figure}
\begin{center}
\includegraphics[width=0.48\columnwidth]{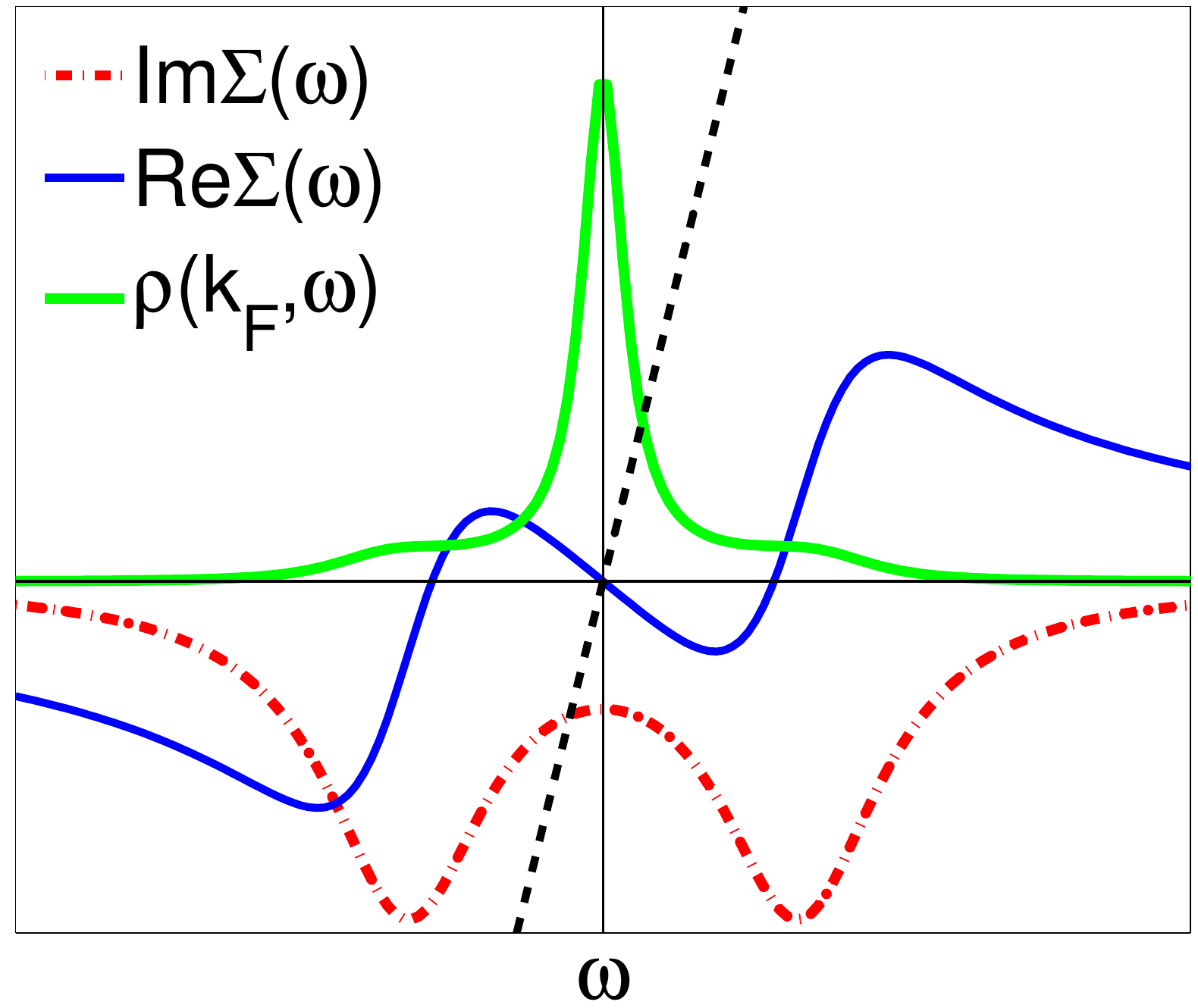}
\hspace{0.1cm}
\includegraphics[width=0.48\columnwidth]{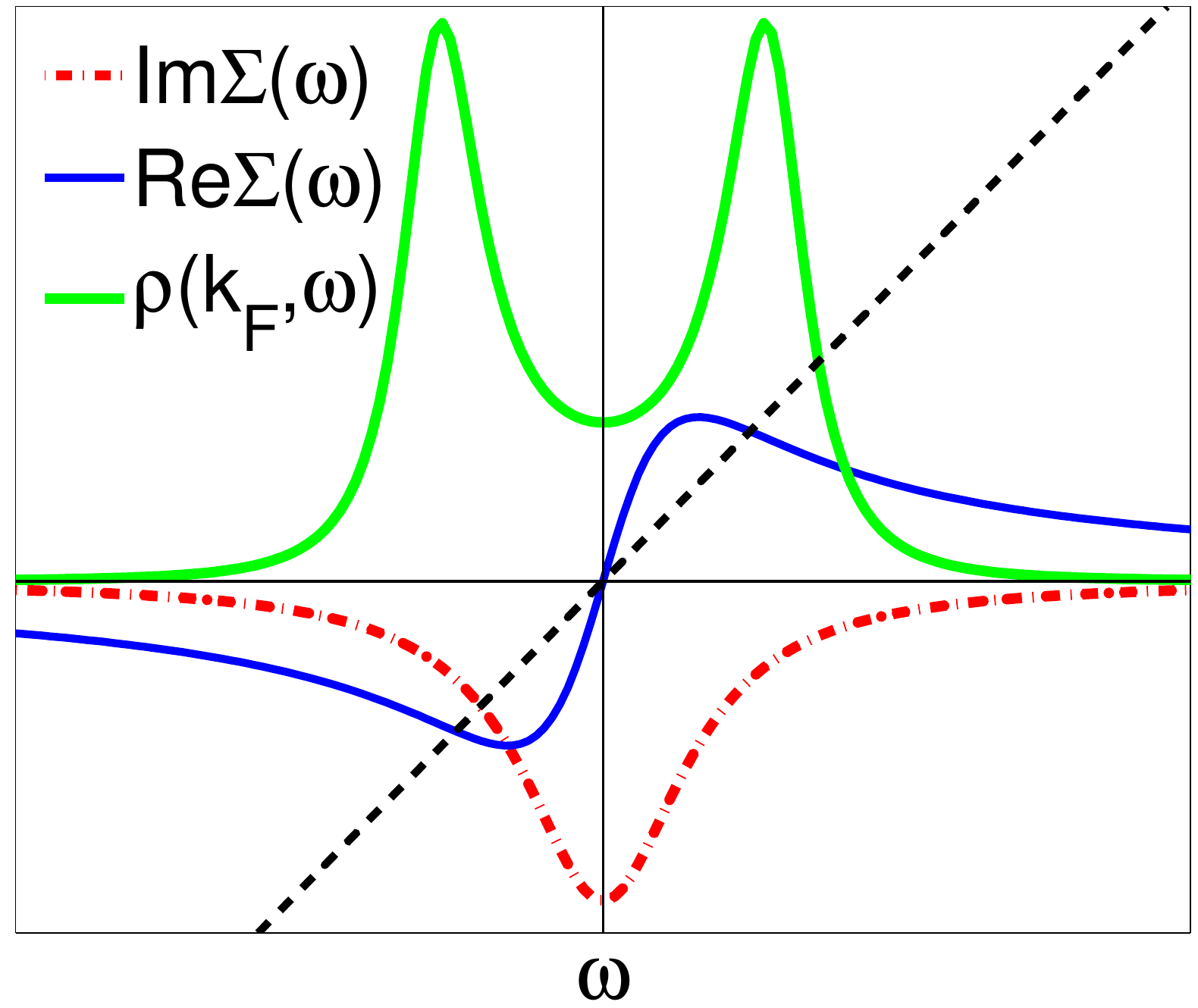}
\vspace{-1cm}
\end{center}
\caption{(Color online) Schematic plot for real and imaginary part of the
  self-energy $\Sigma(\omega)$ in the FL (left) and NFL (right) regime. We
  also show the corresponding spectral function $\rho(\vkF,\omega)$ which
  shows PG behavior in the NFL regime. The dashed diagonal line, $\omega$,
  helps to identify the solutions of the equation
  $\omega=\Real\Sigma(\omega)$, and those positions roughly coincide with
  the PG peaks.} 
\label{fig:sigma_schematic}
\end{figure}

We now discuss the occurrence of a PG for momenta close to the Fermi
surface in the situation where no off-diagonal self-energy is present. Thus
consider $\vk=\vkF$ (interacting Fermi surface) such that 
$\xi_{\vkF}-\Sigma^R(0)=0$.\cite{fn2}
Then we can write
\begin{equation}
  \rho_{\vkF}(\omega)=-\frac{1}{\pi}\frac{\Sigma^I(\omega)}{[\omega-\overline{\Sigma}^R(\omega)]^2+\Sigma^I(\omega)^2},
\end{equation}
where $\overline{\Sigma}^R(\omega)=\Sigma^R(\omega)-\Sigma^R(0)$. 
Provided that $\Sigma^I(\omega)$ does not vary rapidly, we expect $\rho_{\vkF}(\omega)$ to be peaked when
the implicit equation $\omega=\overline{\Sigma}^R(\omega)$ is
satisfied. According to our definitions there is always a solution to this
equation for $\omega=0$. In a weakly interacting system at low 
temperature $|\Sigma^I(\omega)|$ usually has a local minimum at $\omega=0$,
\begin{equation}
  \label{eq:FL}
\Imag \Sigma(\omega)=-a(T)-b\omega^2,  
\end{equation}
where $a(T)\to 0$ for $T\to 0$ and $a,b>0$.
By the Kramers-Kronig relation $\partial_{\omega}\overline{\Sigma}^R(0)<0$ [see
Fig.~\ref{fig:sigma_schematic} (left)]. Then the only solution
of $\omega=\overline{\Sigma}^R(\omega)$ is the one at $\omega=0$. This is the
Fermi liquid peak in the spectral function at $\omega=0$ with width $\sim z
|\Sigma^I(0)|$ and weight $z$, where $z^{-1}=1-\partial_{\omega}\overline{\Sigma}^R(0)$.   
We define the low energy behavior in Eq.~(\ref{eq:FL}) as Fermi liquid (FL) regime.

A PG is obtained with different behavior.\cite{KS90a,KS90b}
If $|\Sigma^I(\omega)|$ possesses a local maximum at $\omega=0$,
\begin{equation}
  \label{eq:NFL}
 \Imag \Sigma(\omega)=-a(T)+b\omega^2,
\end{equation}
then $\partial_{\omega}\overline{\Sigma}^R(0)>0$. If the slope is large enough we
will then encounter additional solutions of
$\omega=\overline{\Sigma}^R(\omega)$ as can be easily seen 
graphically [see Fig.~\ref{fig:sigma_schematic} (right)]. Whether this is the
case depends on the interaction strength, filling fraction and
temperature. Since $|\Sigma^I(\omega)|$ is decreasing, we obtain a local 
minimum at $\omega=0$ in the spectral function and broadened peaks at finite
energies. 
This means that the original peak at $\omega=0$ is split and hence
we obtain a PG. Notice that a local maximum of $|\Sigma^I(\omega)|$
does not necessarily lead to a PG, if the self-energy is not large enough.
In the following we call the low energy behavior of Eq.~(\ref{eq:NFL})
Non-Fermi liquid (NFL) behavior.    
As we have discussed above $|\Sigma^I(\omega)|$ is typically maximal at
$\omega=0$ at high temperature when the physics becomes dominated by local
interactions. It is also directly visible in the phase space factor appearing
in the second order perturbation theory in $U$ (see appendix). 
Therefore, at high temperature we expect NFL behavior, and at low
temperature we usually have FL behavior.
We define the crossover scale as $T_{\rm FL}$, i.e., where the behavior of
$\Sigma^I(\omega)$ changes from Eq.~(\ref{eq:NFL}) to (\ref{eq:FL}). 
In this picture PG behavior in $\rho_{\vkF}(\omega)$ occurs therefore as long
as (i) $U$ is large enough ($\sim W$) and (ii) $T> T_{\rm FL}(U)$. In
particular, the PG is always present above $T_c$ if $T_c>T_{\rm FL}(U)$.

\section{PG physics at half filling} 
In this section we analyze results from the DMFT calculations for spectral
functions and self-energies and 
focus on the situation at half filling.
An overview of the different regimes as function of $U$ and $T$ is shown
in Fig.~\ref{phase_diagram}.   

The phase diagram includes the SF phase and the regimes where the self-energy shows FL and
NFL behavior as defined in Eq.~(\ref{eq:FL}) and Eq.~(\ref{eq:NFL}),
respectively. By performing calculations suppressing the SF phase below $T_c$,
we find that the boundary between FL and NFL regimes (not shown) is connected
to the bipolaron transition at $T=0$, which is equivalent to the Mott
transition for repulsive interactions.   
The NFL regime in the phase diagram is separated into a region for stronger
interactions where we observe a PG in the integrated spectral function, and a region without PG (no PG) for weaker
couplings. 

\begin{figure}[tb]
\begin{center}
\includegraphics[width=0.95\columnwidth]{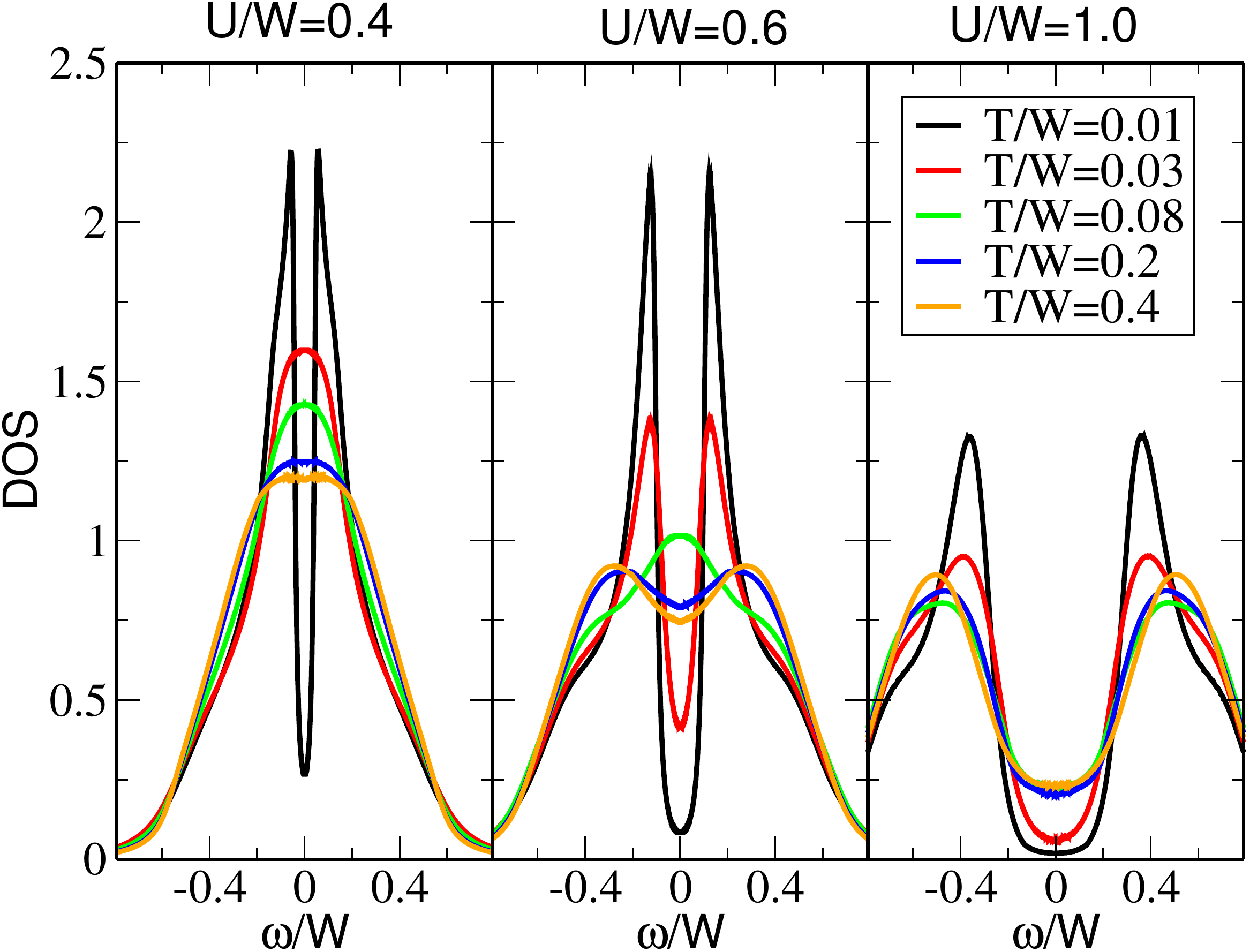}
\includegraphics[width=0.95\columnwidth]{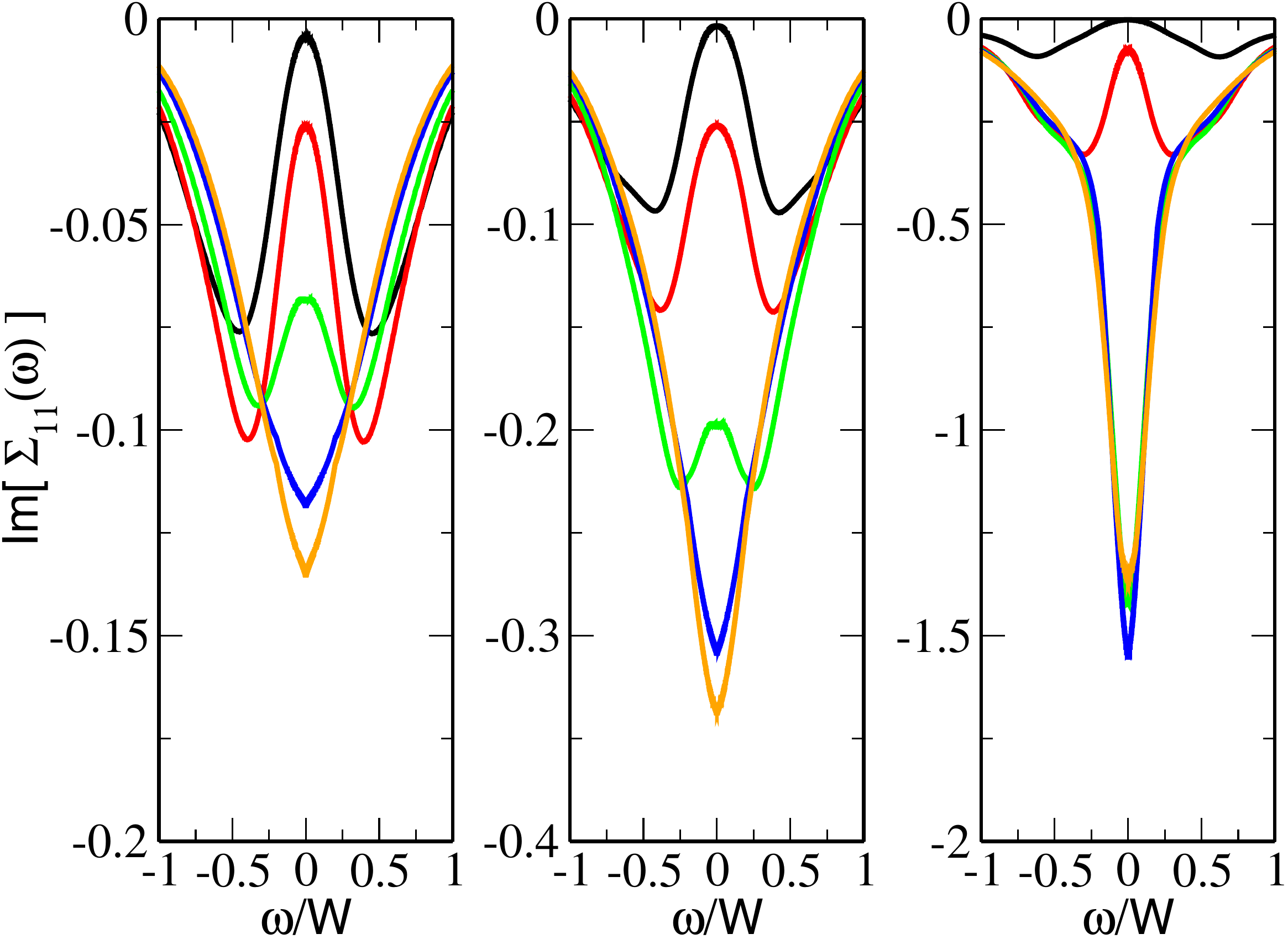}
\end{center}
\caption{(Color online) Integrated DOS, $\rho_{11}(\omega)$, and imaginary
  part of the self-energy, $\Sigma_{11}(\omega)$, for different interaction
  strengths and temperatures.\label{compare}}   
\end{figure}
In the upper panels of Fig.~\ref{compare} we show the interacting local
DOS $\rho(\omega)$. At weak coupling and intermediate temperatures, $\rho(\omega)$ very much
resembles the non-interaction DOS, $\rho_0(\omega)$, and the small
self-energy does not have a pronounced effect. 
Although $|\Imag\Sigma|$ is peaked at the Fermi energy for $U=0.4W$ at high
temperatures, there is no PG structure in the DOS.  
In contrast, for larger interactions, $U/W=0.6$, $U/W=1$, we find at high
temperatures a PG structure of two peaks at $\pm U/2$ and
a suppression of the density of states at $\omega=0$. The behavior
is more pronounced for larger interactions. In both cases the magnitude of the
PG is clearly related to $U$.
This structure is induced by the NFL peak in the $|\Imag\Sigma|$ (lower
panels). As discussed in the previous section III this result 
can be understood in terms of the local excitations dominating the physics at
high temperature. 
At weak and intermediate interaction strengths the system crosses
over to a FL regime before $T_c$ is reached when decreasing the
temperature. For $U/W=0.4$ and $U/W=0.6$, $|\Imag\Sigma(\omega)|$ exhibits a dip at
the Fermi energy at low enough temperature, 
$T/W<0.1$, which is accompanied with a peak structure in the DOS. Such a
change in the behavior of self-energy and DOS cannot be observed for strong
coupling, where the PG structure exists for all temperatures above $T_c$. At
very low temperatures, the system is in the SF 
phase in all cases, which is characterized by a gap in the DOS, which
coincides with a dip in $\Imag\Sigma(\omega)$.  
So even though the two cases, $U/W=0.6$, $U/W=1$, in Fig.~\ref{compare} look similar at high
temperature (PG) and very low temperature (SF gap), they display
a striking difference for intermediate temperatures. For the larger
coupling strength the SF transition occurs from a PG state (see also
Fig.~\ref{TC}); in contrast for $U/W=0.6$, the SF instability happens in the
FL regime.

\begin{figure}[tb]
\begin{center}
\includegraphics[width=0.95\columnwidth]{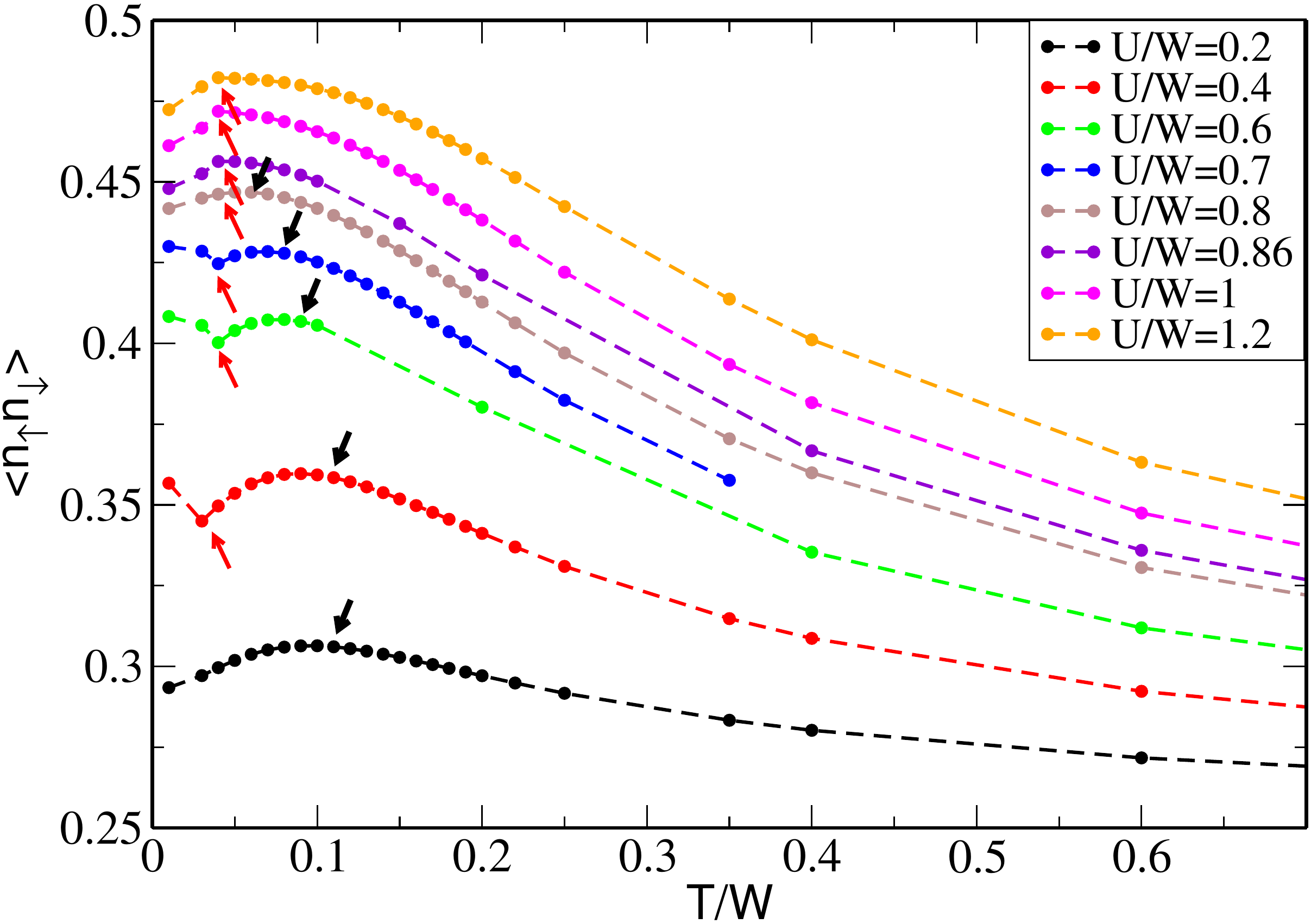}
\end{center}
\caption{(Color online) The local pair density $\langle n_\uparrow n_\downarrow\rangle$
  for different temperatures and interaction strengths. The black arrow marks
  the transition from non-Fermi-liquid to Fermi-liquid behavior. The red
  arrow marks the transition into the SF phase. \label{double}}
\end{figure}

\begin{figure}[tb]
\begin{center}
\includegraphics[width=0.95\columnwidth]{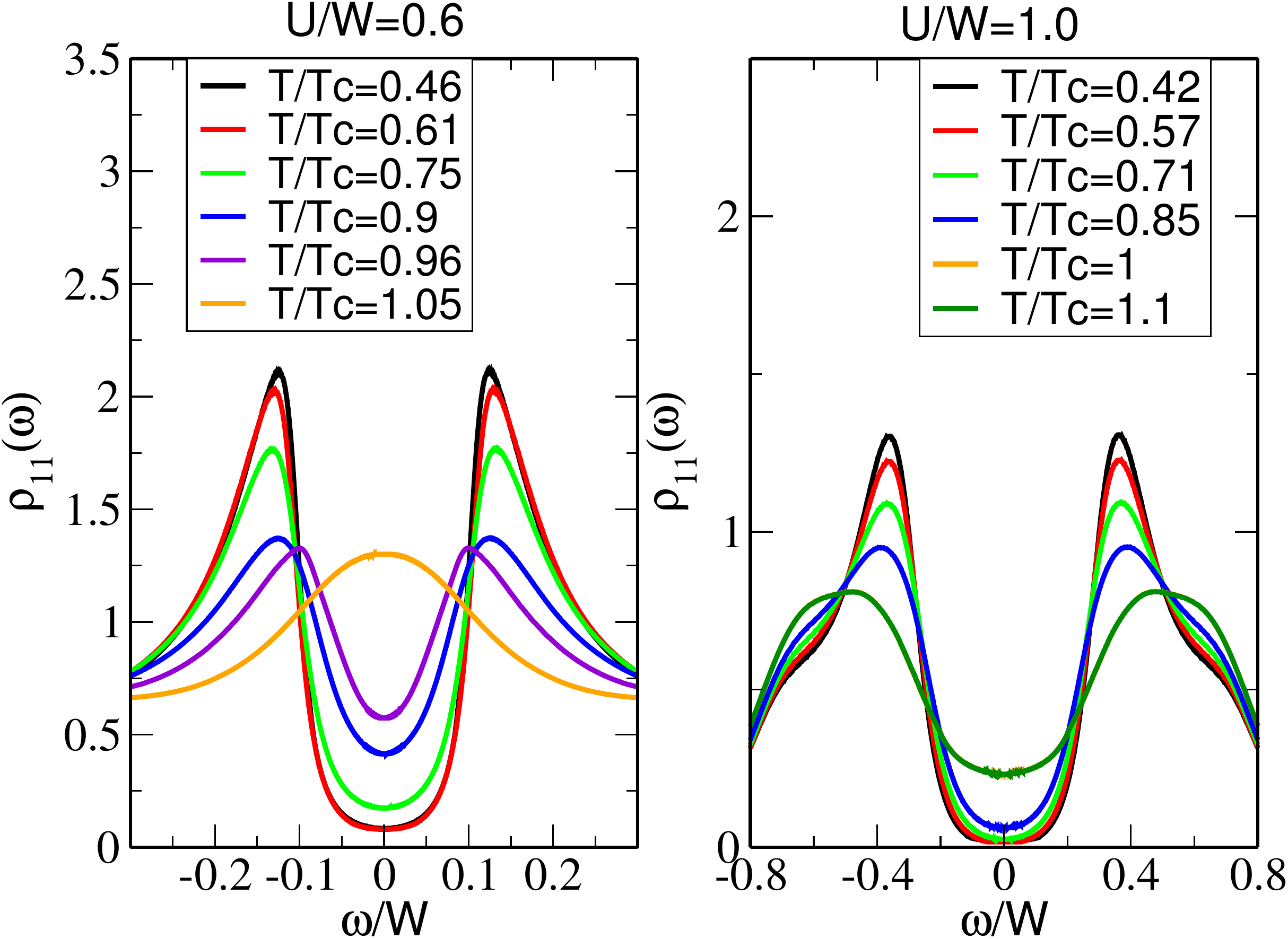}
\includegraphics[width=0.95\columnwidth]{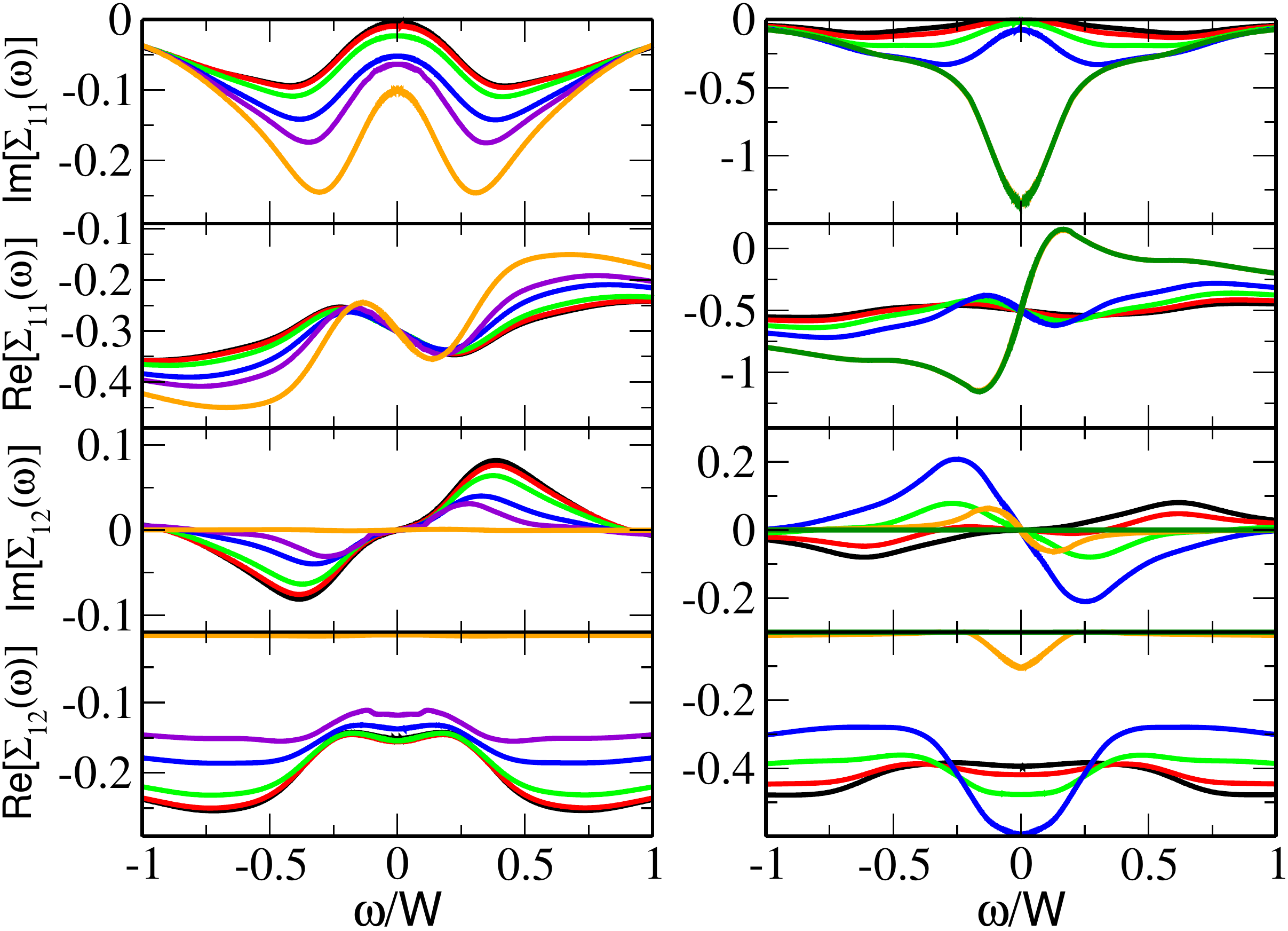}
\end{center}
\caption{(Color online) $\rho(\omega)$ and $\Sigma(\omega)$ close to the
  SF phase transition. We use the same legend for the self-energies as
  for the Greens functions. The transition temperatures are $T_c/W=0.32$ for
  $U/W=0.6$ and $T_c/W=0.35$ for $U/W=1.0$.\label{TC}} 
\end{figure}

Further insights can be obtained by studying the behavior of the double
occupancy or local pair density, $\langle n_\uparrow n_\downarrow\rangle$,
which is shown in Fig.~\ref{double} for different temperatures and interaction
strengths.

Independent of the interaction strength, in the high temperature limit, $T\gg
W,U$, the pair densities approach the non-interacting values $n_{\sigma}^2$, where
$n_{\sigma}$ is the density for one spin component, at half filling $\langle n_\uparrow
n_\downarrow\rangle=0.25$. 
In the atomic limit, $t=0$, the double occupancy can be easily calculated. 
At half filling, all atomic states are occupied with equal probability, so
that the double occupancy reads 
\begin{equation}
\langle n_\uparrow n_\downarrow\rangle=\frac{1}{2+2\exp(-U/(2T))}.
\end{equation} 
At high temperature, $T/W>0.5$, this formula agrees very well with the results in
Fig.~\ref{double}, demonstrating again that the physics at high temperature
is dominated by local processes. 

Decreasing the temperature,  $\langle n_\uparrow n_\downarrow\rangle$
increases due to the attractive interaction. This 
effect is stronger for stronger interaction. For interaction strengths $U/W<0.8$, we
find a maximum {\em before} the system enters the
SF phase at $T_c$. This maximum appears to be correlated with the crossover
temperature $T_{\rm FL}$ (black arrows) between FL and NFL behavior in the
self-energy.  The disappearance of the maximum in the pair density for
interaction strengths $U/W>0.8$ agrees with the vanishing of the FL regime
phase in the phase diagram. For $U/W<0.8$, the pair density 
decreases when lowering the temperature below $T_{\rm FL}$, but then increases again when entering the SF
phase (arrow at $T_c$). For strong interactions ($U/W>0.8$) on the other hand,
the pair density increases with decreasing temperature until $T_c$ is
reached and then decreases. This agrees with the
known fact that the superfluidity is driven by interaction
energy gain for weak coupling, as opposed to kinetic energy gain for
strongly coupled systems.\cite{TCC05}

With these insights we can comment on how our results compare to the preformed
pair scenario in Fig.~\ref{schemphase_diagram}. 
It is interesting to note that for very high temperatures the PG 
behavior in Fig.~\ref{compare} does not change significantly anymore. In other
words the PG persists and no $T_{\rm p}$ for its appearance can be identified. 
This is the case even for temperatures where the pair density has decreased to values
close to the non-interacting result. 
Furthermore we found PG behavior for the two cases $U/W=0.6$, $U/W=1$ at high
temperature, but for intermediate temperatures ($T/W \sim 0.05$) the case
$U/W=0.6$ shows FL behavior. In both cases we observe a strongly enhanced
local pair density for such temperatures, which can be interpreted as a preformed  pair
state, however, the manifestation in the spectral function is different. 
Both of these observations are in clear contrast to the preformed pair
scenario, where the existence of the PG behavior is linked to the presence of an enhanced pair
density.\cite{Ran10,CLS06}

In Fig.~\ref{TC}, we take a closer look at dynamic response functions close to the
SF transition temperature $T_c$.  

\begin{figure*}[!ht]
\includegraphics[width=0.32\linewidth]{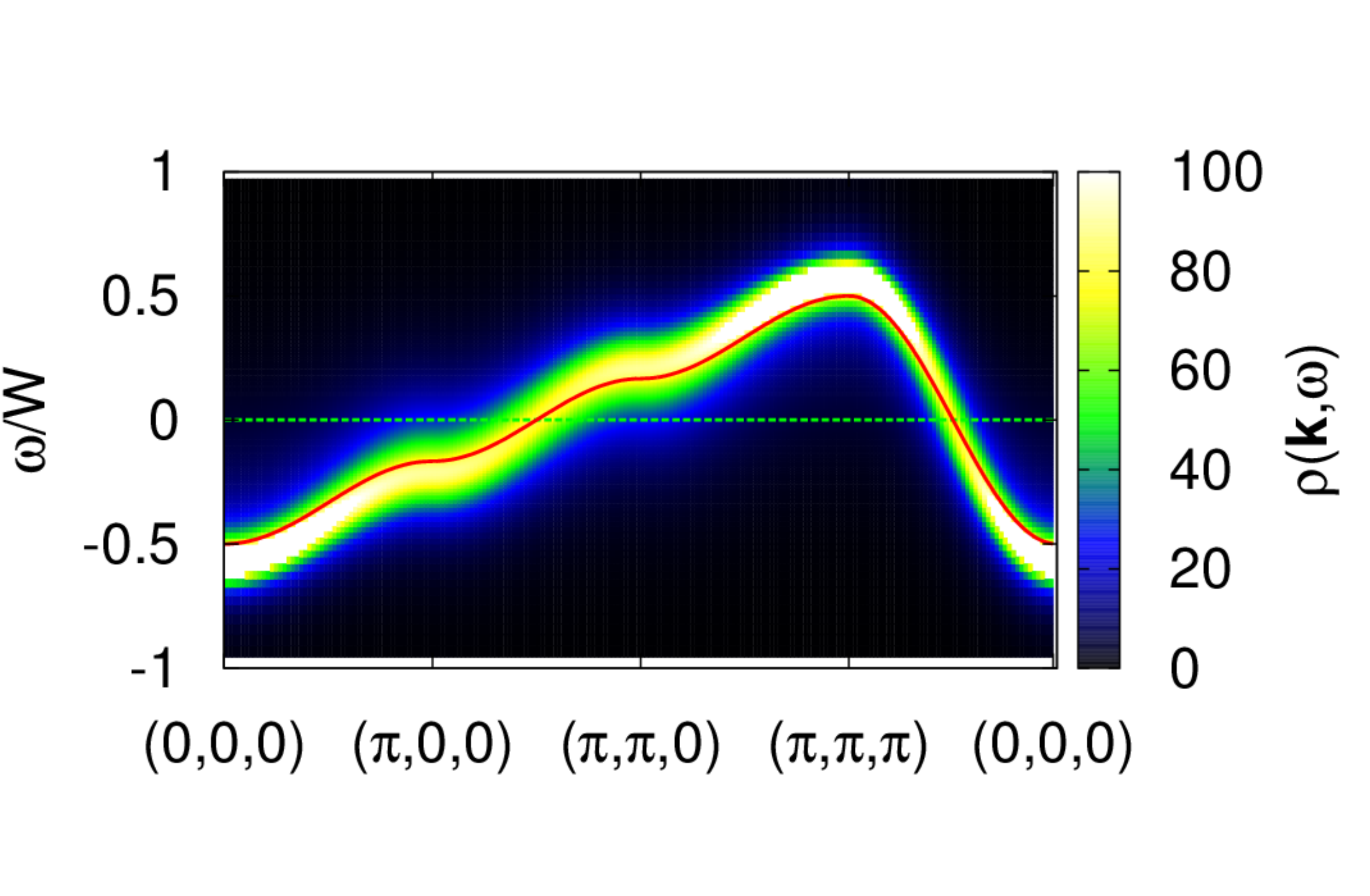}
\includegraphics[width=0.32\linewidth]{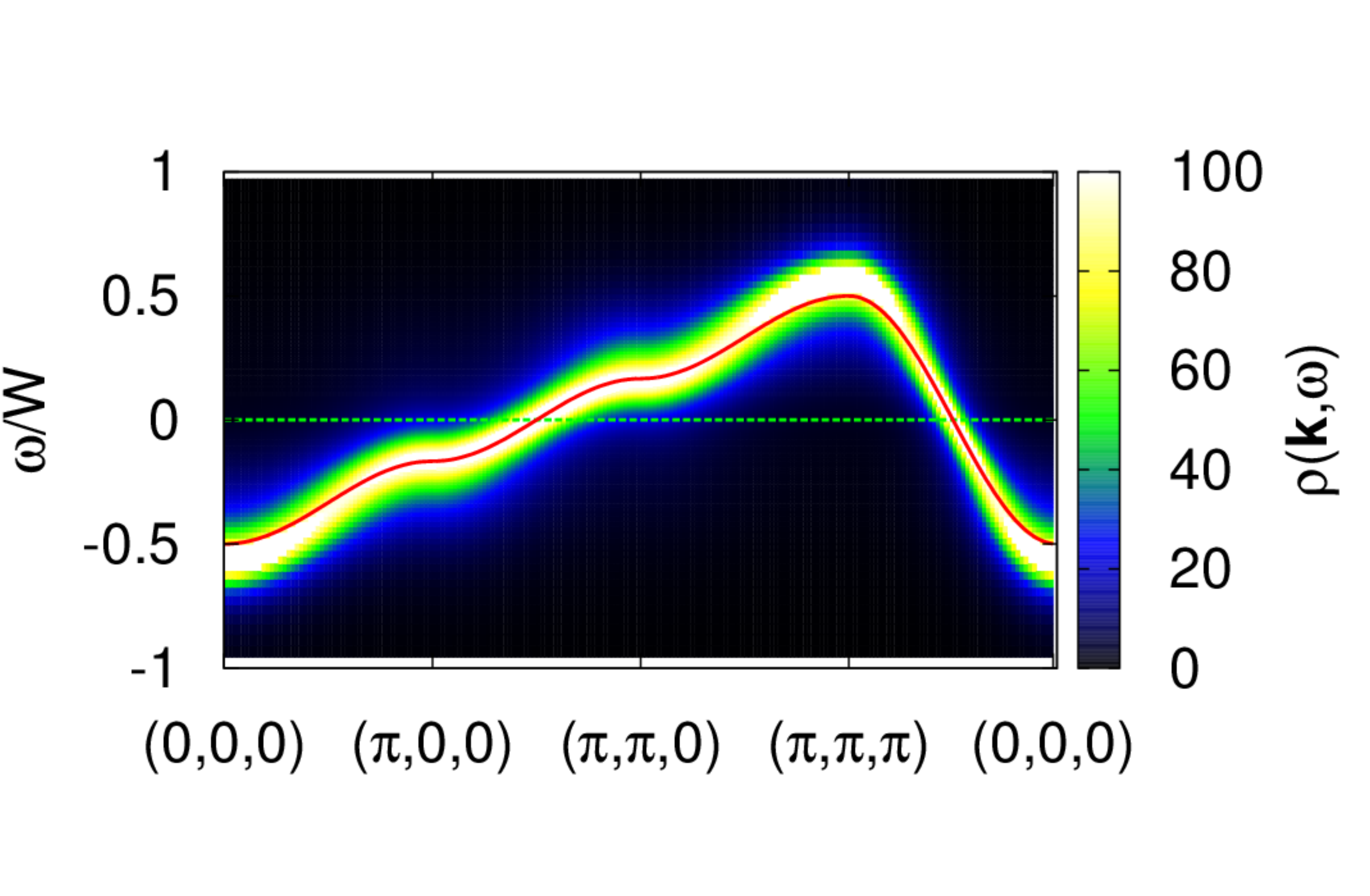}
\includegraphics[width=0.32\linewidth]{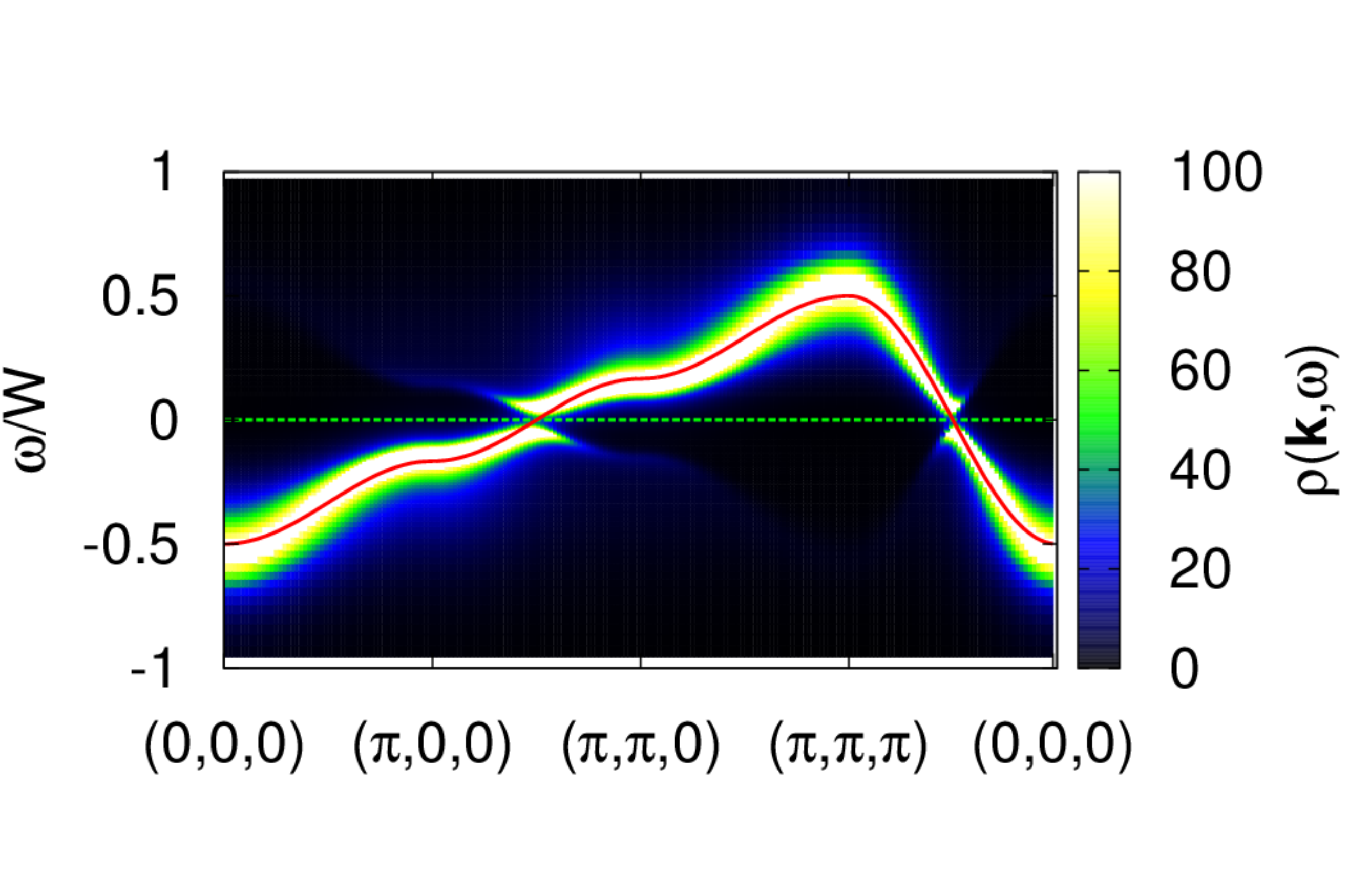}

\includegraphics[width=0.32\linewidth]{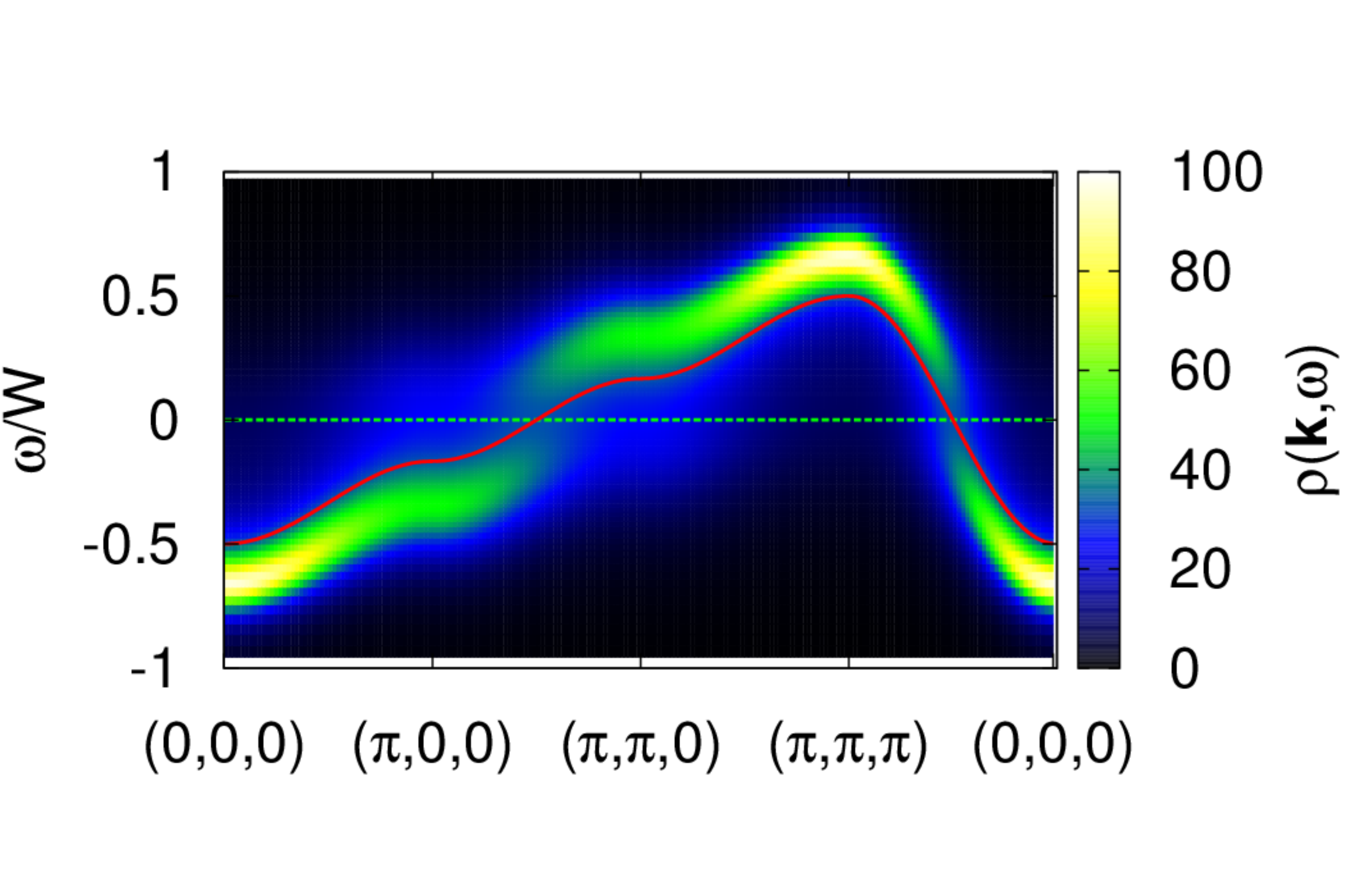}
\includegraphics[width=0.32\linewidth]{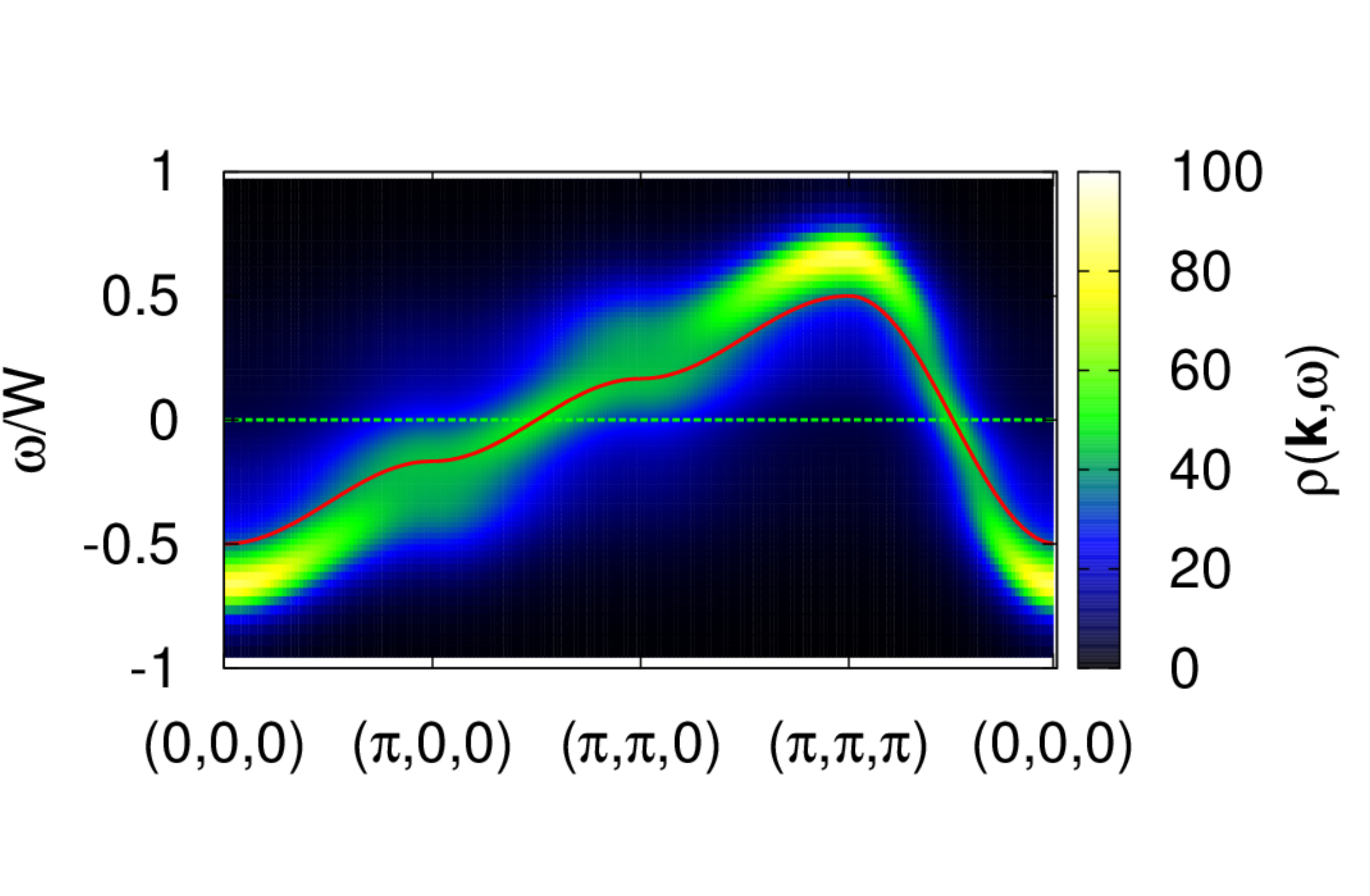}
\includegraphics[width=0.32\linewidth]{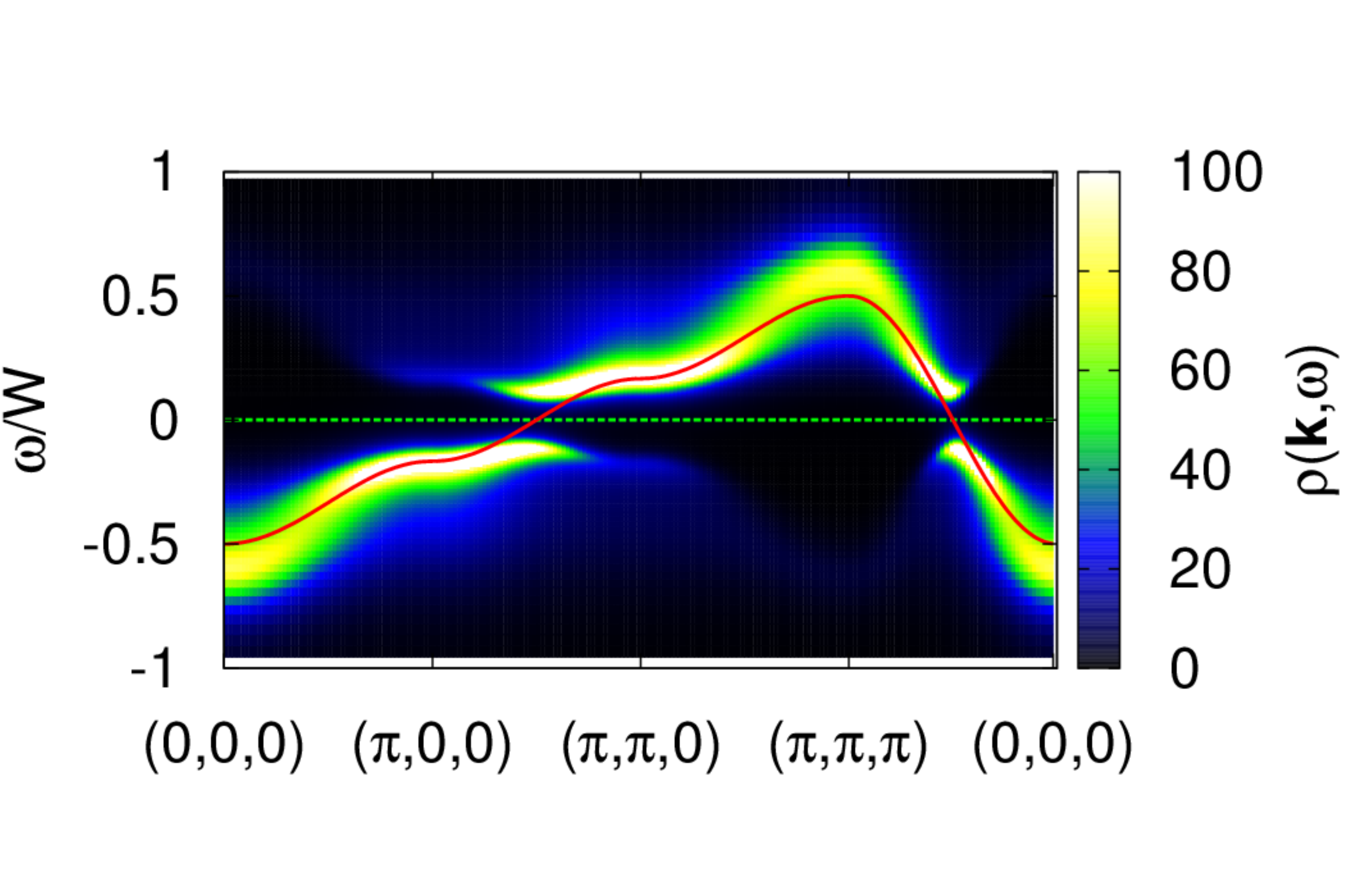}

\includegraphics[width=0.32\linewidth]{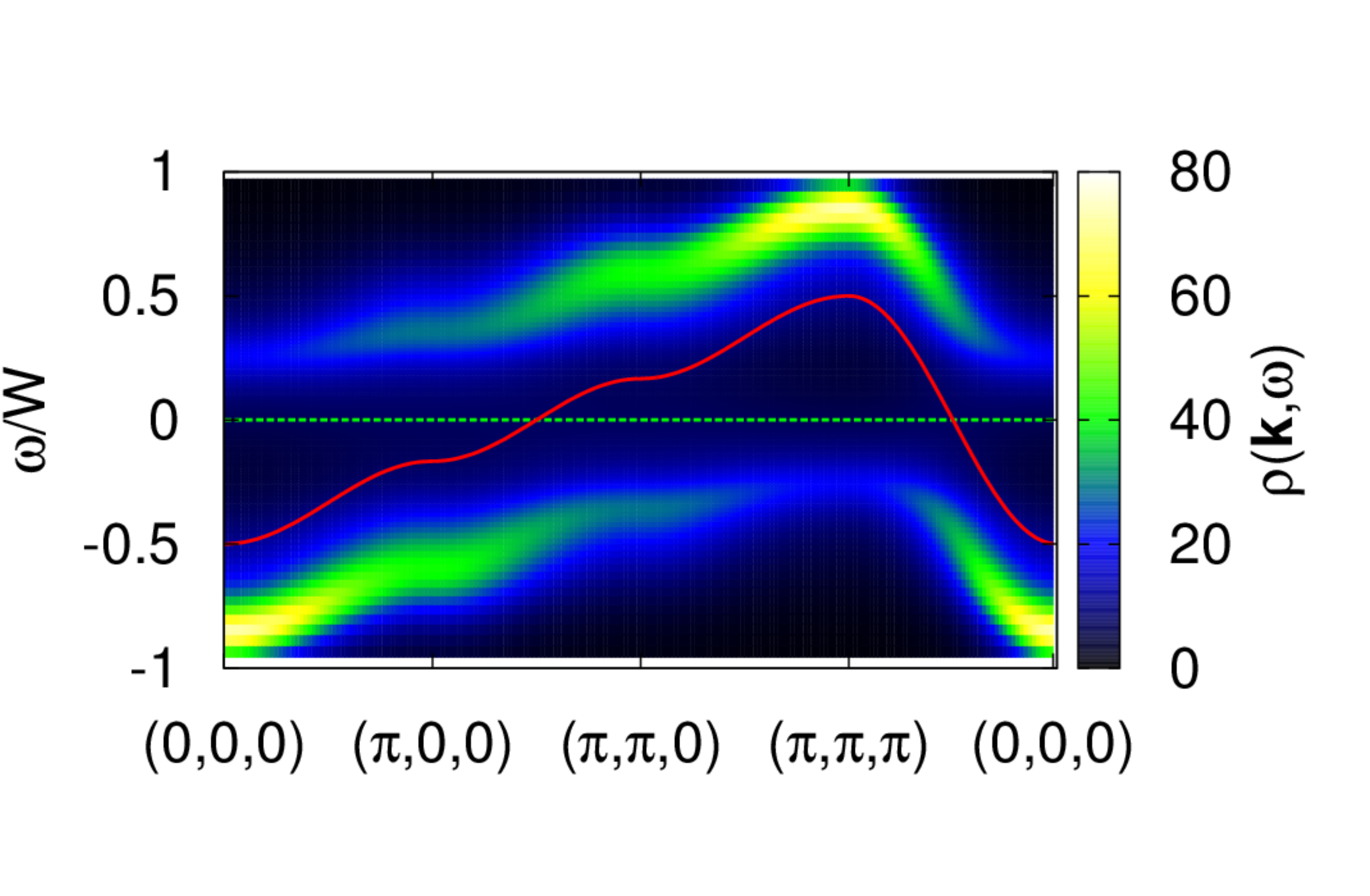}
\includegraphics[width=0.32\linewidth]{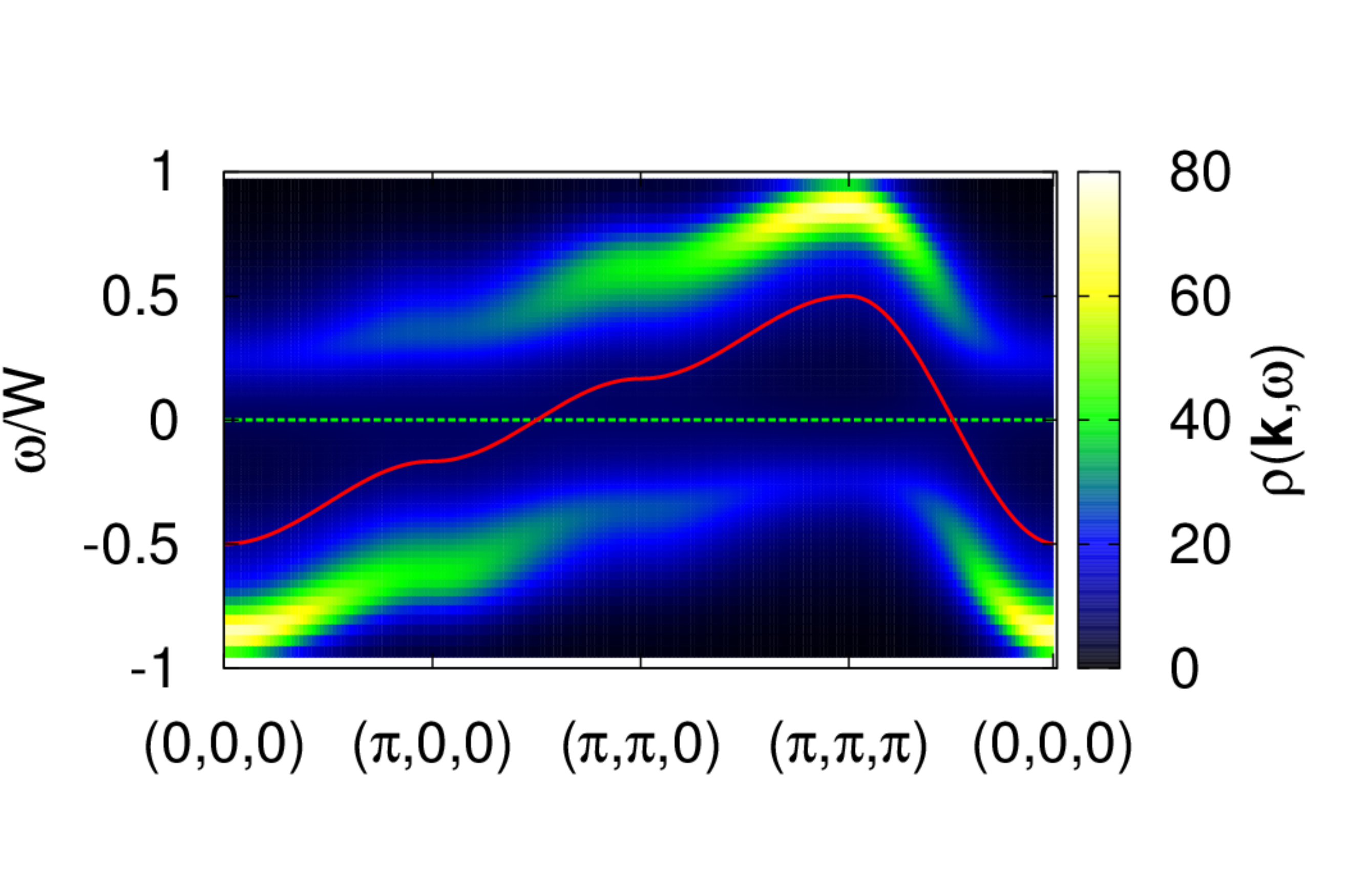}
\includegraphics[width=0.32\linewidth]{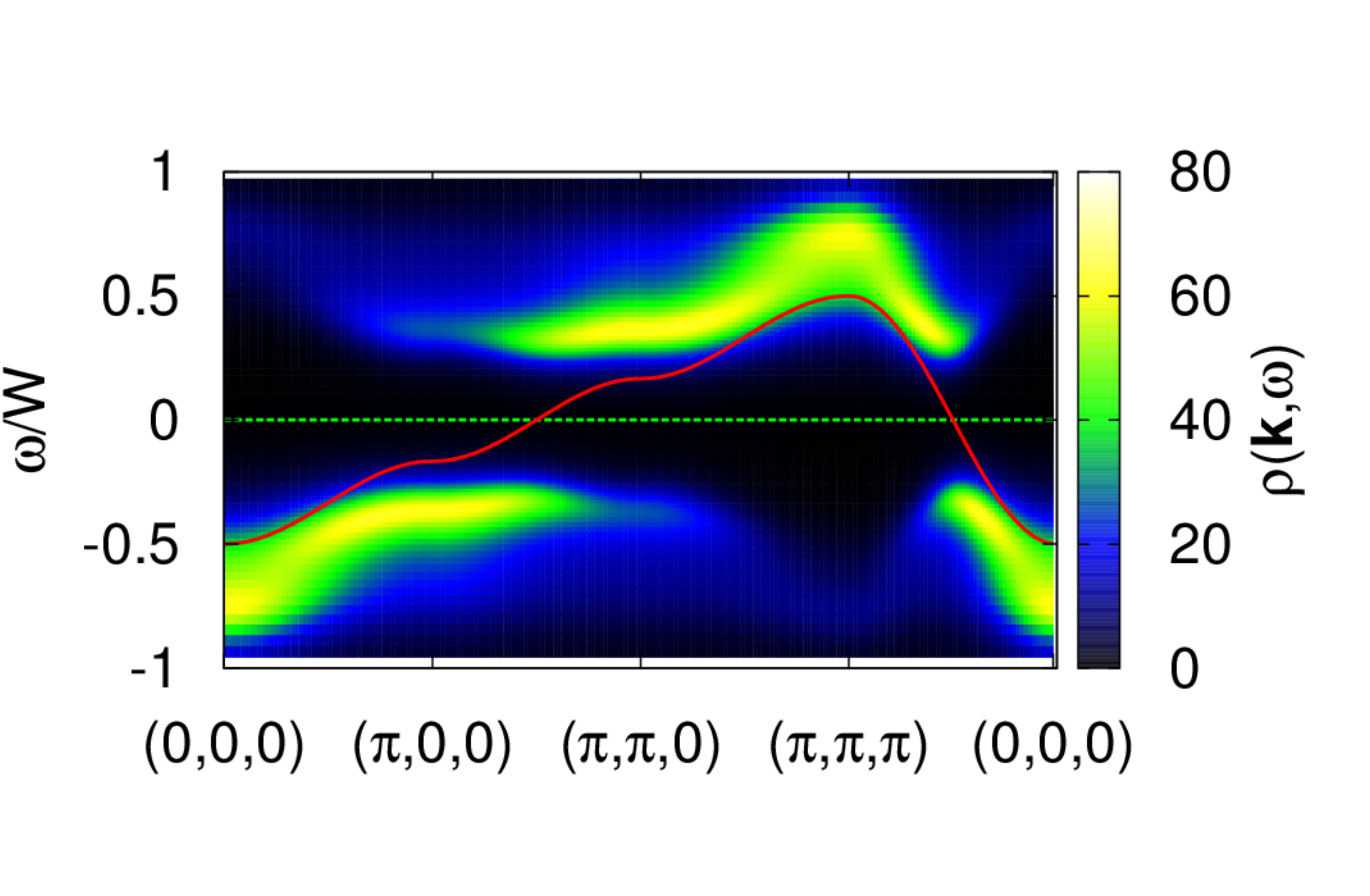}

\caption{(Color online) Momentum resolved spectral
  function for $U/W=0.4$ (upper panel), $U/W=0.6$ (middle panels) and $U/W=1$ (lower panels). The
  temperature are $T/W=0.2, 0.08, 0.01$ from left to right. The red line
  corresponds to the non-interacting system, the green line corresponds to the
  Fermi energy.\label{spec_momentum}} 
\end{figure*}

The plots in the upper part of figure show $\rho(\omega)$ for $U/W=0.6$ 
and $U/W=1$, which correspond to a transition into the SF phase from the FL
and NFL regime, respectively. The lower part of the figure
displays the corresponding diagonal and off-diagonal self-energies (real and
imaginary parts). For the weaker coupling case, $U/W=0.6$, at $T>T_c$
there is the usual FL dip in $\Imag\Sigma_{11}(\omega)$ and the corresponding peak in $\rho(\omega)$.  
When the temperature is lowered through $T_c$ the off-diagonal self-energy
becomes finite and a dip in $\rho(\omega)$ is induced.
Very close below the transition temperature, the main effects for this come
from $\Real\Sigma_{12}(\omega)$. Lowering the temperature further, the
amplitude of the diagonal part of the self-energies decreases without showing
new features.  
As discussed in Sec.~III the gapping out of excitation is dominated by
contributions from $\Real\Sigma_{12}(\omega)$.

In the case of stronger interaction, $U/W=1$, superfluidity sets in
the NFL regime with a PG at the Fermi energy. 
When lowering the temperature through $T_c$, the off-diagonal
self-energy becomes finite, but at first the diagonal part of the self-energy
remains nearly unchanged (The orange line, $T/T_c=1$, overlaps with the dark
green line, $T/T_c=1.1$).   
On further reducing the temperature the off-diagonal self-energy increases
substantially and $\Imag\Sigma_{11}(\omega)$ is strongly reduced developing a FL dip at
the Fermi energy. 
The gap in $\rho(\omega)$ changes smoothly from the PG with broad peaks separated by $U$
to the sharper structures (coherence peaks) in the
SF phase. It is interesting to note that the gap, if defined as
the distance between the maxima is larger above $T_c$ in the PG regime than in the SF
phase. One should also note that for low temperatures the gap becomes much
more pronounced with a suppression of spectral weight at $\omega=0$ and as such
is approaching a full gap in the limit $T\to 0$. 

A remarkable observation is that the qualitative behavior of the off-diagonal
part of the self-energy can change within the SF phase. 
Generally, $|\Real\Sigma_{12}(\omega)|$ approaches the mean field result $U\langle
c_{i,\uparrow} c_{i,\downarrow}\rangle$ for $|\omega|\to \infty$.\cite{BHD09}
At weaker coupling ($U/W=0.6$) and low temperature it is minimal for small $\omega$.
Decreasing the temperature, the anomalous expectation value increases and this
is reflected in the results for $\Sigma_{12}(\omega)$. The $\omega$-dependence
can be understood at weak coupling from the effective interaction for inducing 
superfluidity, which possesses a repulsive component which is peaked for small
$\omega$.\cite{fn3}
However, when entering the SF phase from the PG regime at
stronger coupling ($U/W=1$), $|\Real \Sigma_{12}(\omega)|$ first develops a
strong maximum at $\omega=0$. When the temperature is lowered further this
behavior continually reverts to the one of the weak coupling situation. The form of the
spectral function changes at $T_c$ and there is a shift from the gap feature
being induced by $\Sigma_{11}(\omega)$ (above $T_c$) to $\Sigma_{12}(\omega)$
(below $T_c$). The observed strong changes are related to this shift and
a more thorough understanding requires further investigation.

We now turn our attention to features in the momentum resolved spectral function
$\rho_{\vk}(\omega)$. A good overview of the behavior for different
interactions and temperatures can be obtained in the intensity plots in
Fig.~\ref{spec_momentum}. 

\begin{figure}[tb]
\begin{center}
\includegraphics[width=0.95\columnwidth]{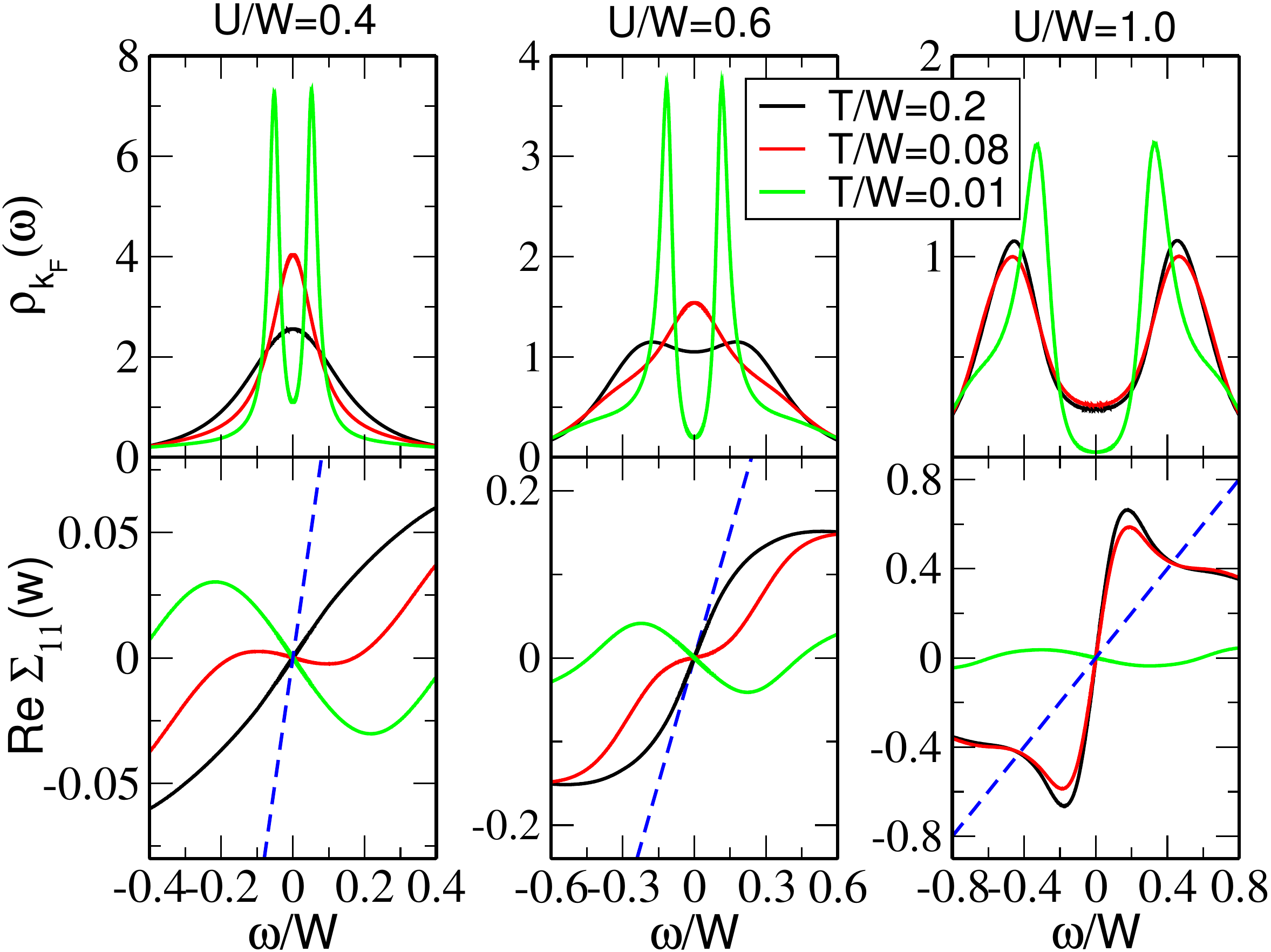}
\end{center}
\caption{(Color online) $\rho_{\vkF}(\omega)$ for $\vkF=(\pi/2,\pi/2,\pi/2)$ for $U/W=0.4$,
  $U/W=0.6$ and $U/W=1$ for different temperatures.\label{compare2}}  
\end{figure}

We show $\rho_{\vk}(\omega)$ for three
interaction strengths [$U/W=0.4$ (upper panels), $U/W=0.6$ (middle panels),
and $U/W=1$ (lower panels)] for three different temperatures [$T/W=0.2$ (left), $T/W=0.08$ (middle), and
$T/W=0.01$ (right)]. We also show the Fermi level (dashed line) and the
non-interacting dispersion (full red line) as an orientation. 
At weak coupling, $U/W=0.4$, the spectral function only displays a weak
modification from the non-interacting result with certain broadening of the
peaks and a minor shift of spectral weight. At low temperature the system is
SF and excitations at $\vkF$ are gapped out. Notice that the width of the 
Bogoliubov peaks at the gap edge is overestimated by our broadening
procedure.\cite{BHD09}

For $U/W=0.6$, we find similar features for $\rho_{\vk}(\omega)$ as what has
been found for the integrated spectral function, $\rho(\omega)$, as far as the PG is concerned. At high
temperatures we see a broadened dispersion similar but shifted from the
non-interacting one. Spectral weight is suppressed at the Fermi energy such
that PG features are realized at high temperatures. Curiously, this PG closes at
intermediated temperatures $\sim T_{\rm FL}$ where the behavior of the
self-energy changes. Below $T_c$ the spectrum is gapped again. Notice that
band renormalization features appear somewhat weaker than at high
temperatures. 

For $U/W=1$ the NFL regime extends from high temperatures down to
$T_c$. The self-energy undergoes only very slight changes
when decreasing the temperature in the NFL regime. Accordingly, the
momentum-resolved spectral function for $T/W=0.2$ and $T/W=0.08$
(lower left panel and lower middle 
panel) are nearly the same. We observe a large PG around the
Fermi energy; the spectral weight at the Fermi energy is very small. When
entering the SF phase, gap features are visible and the dispersion changes in
the vicinity of $\omega=0$. For this interaction strength, we observe a clear
deviation between the non-interacting band structure and the interacting
spectral function. 
In the SF phase we find a mirror or ``shadow''
band appearing as reflected from $\omega=0$. 
These bands can be understood due to a particle-hole doubling
in the Nambu representation. This is an effect also observed in the
antiferromagnetically ordered phase with zone doubling.\cite{BH07c}

In Fig.~\ref{compare2} we show particular cuts for $\rho_{\vkF}(\omega)$ as a
function of $\omega$. Here we can see the PG features even more
clearly. Similar as in the integrated spectrum, $\rho(\omega)$, the PG is absent for the weak coupling
case, $U/W=0.4$, but present at high temperature for stronger interactions, $U/W=0.6$ and
$U/W=1$. For $U/W=0.6$ the PG disappears in the FL regime,
whereas it remains for $U/W=1$. We also
show the real part of the diagonal self-energy. As discussed in Sec.~III,
the peak splitting in $\rho_{\vkF}(\omega)$ can be induced from non-trivial
solutions of $\omega=\Real\overline{\Sigma}_{11}(\omega)$, and we have
included a dashed line to see this graphically. As can be clearly seen, this 
condition is not satisfied  in the weak coupling case. In contrast, at strong
coupling, $U/W=1$, the intersection points characterize the peak positions
well. The spectral function changes in the SF phase (lowest temperatures), where the
coherence peaks at the gap edge become visible.

\begin{figure}[tb]
\begin{center}
\includegraphics[width=0.95\columnwidth]{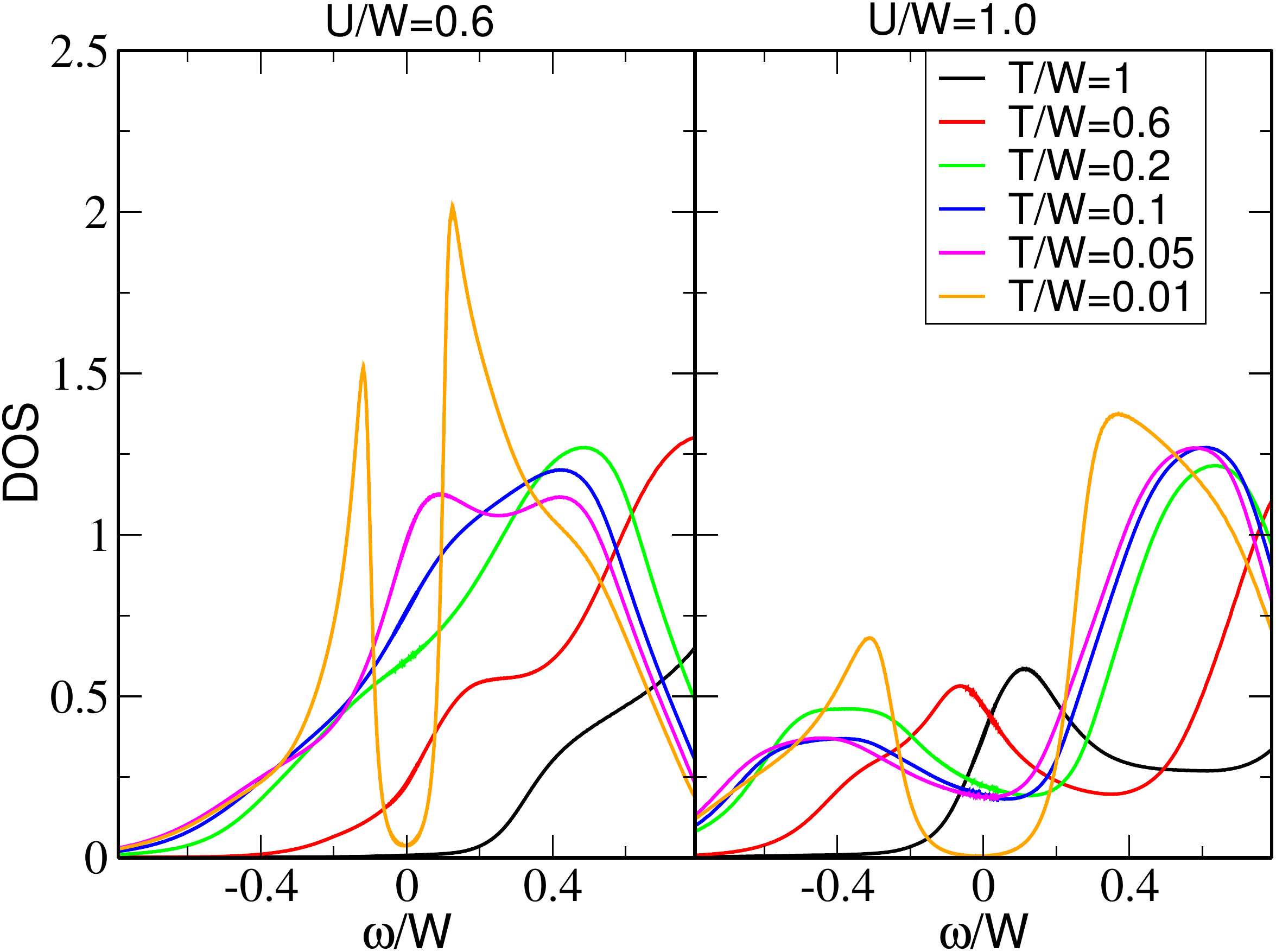}
\includegraphics[width=0.95\columnwidth]{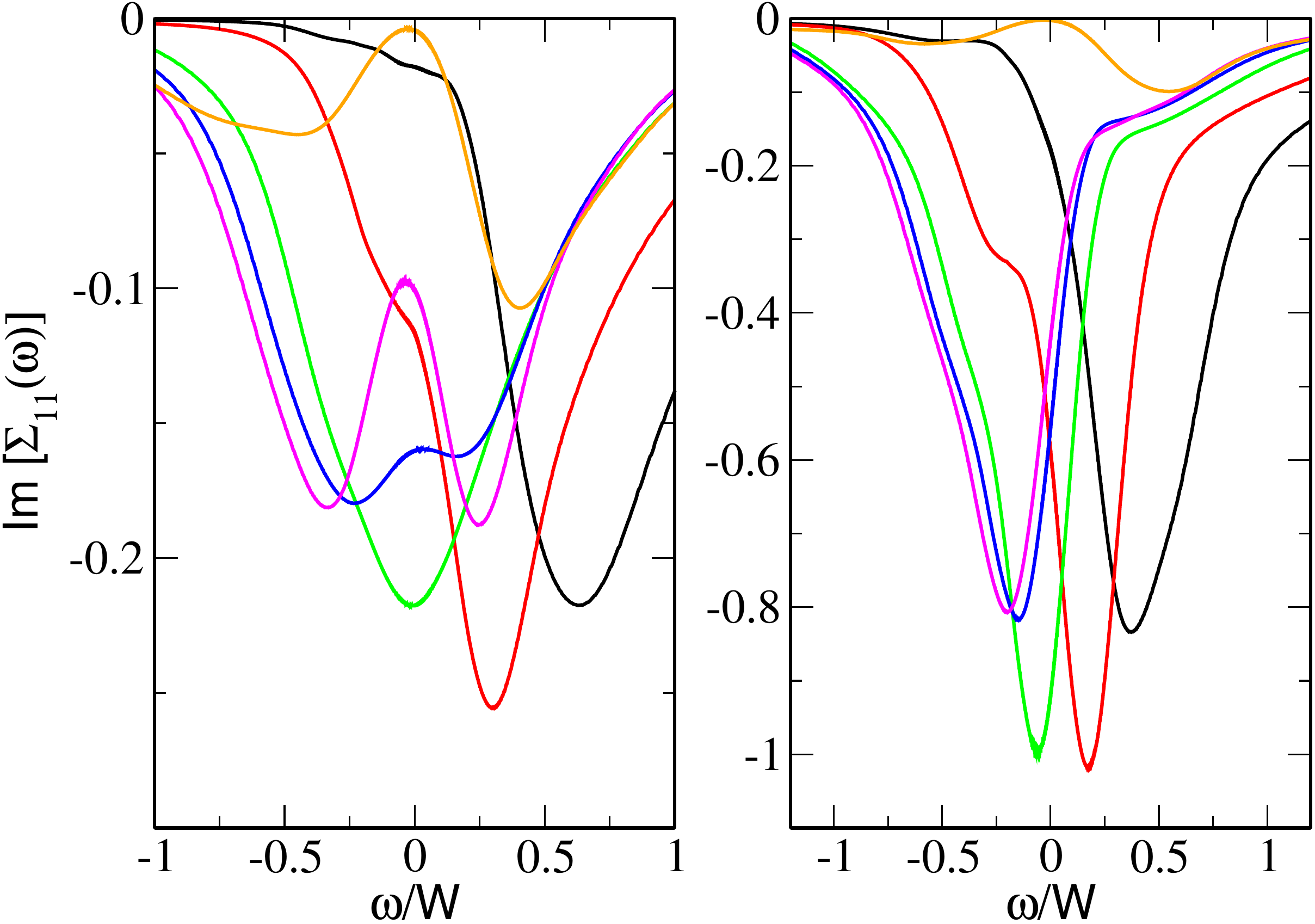}
\end{center}
\caption{(Color online) The DOS and imaginary part of the self-energy for 
  $U/W=0.6$ and $U/W=1$ for different temperatures. The filling of the system
  is fixed to $n=0.5$.\label{compare_doped}}  
\end{figure}

\begin{figure}[tb]
\begin{center}
\includegraphics[width=0.95\columnwidth]{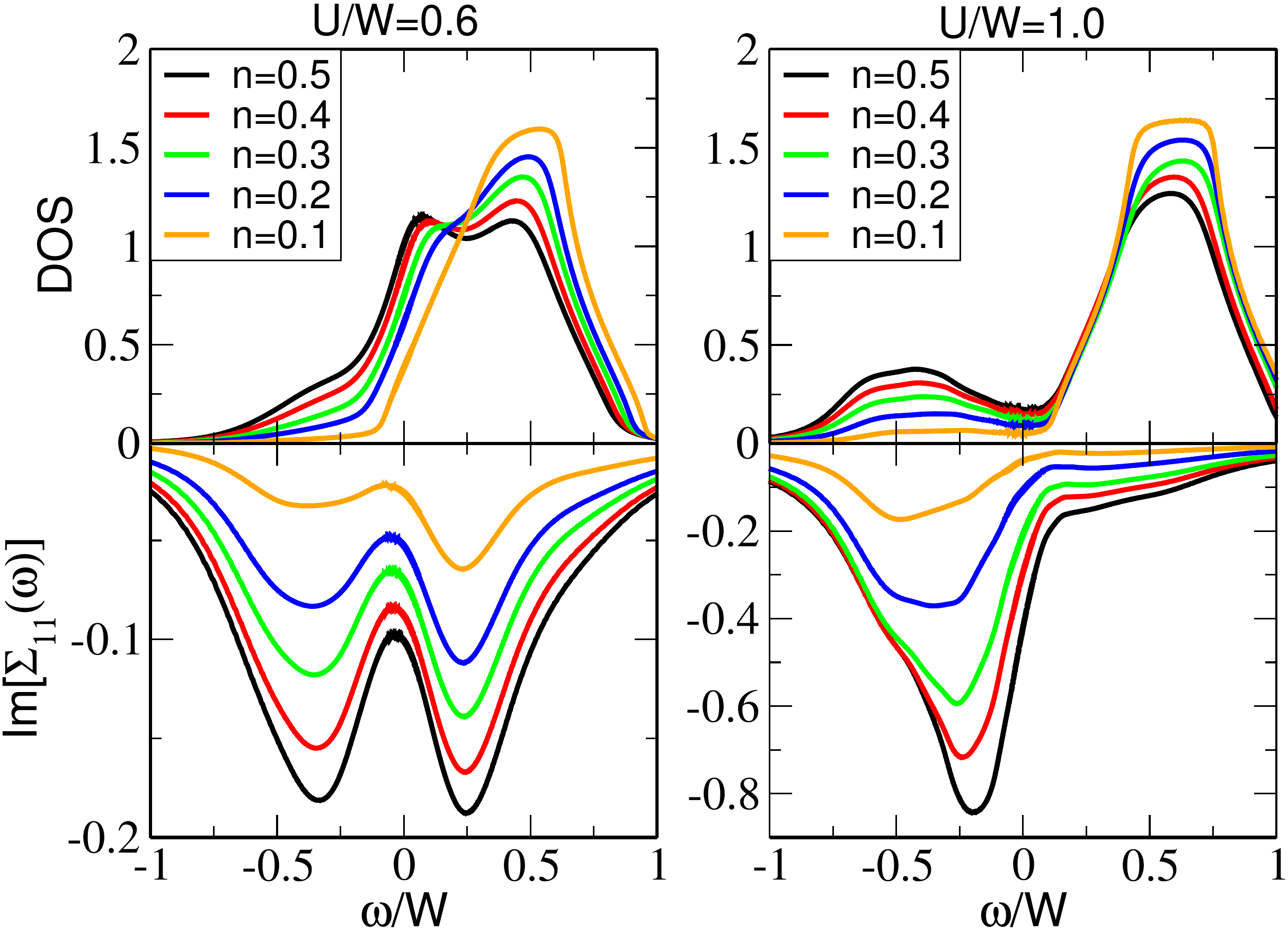}
\end{center}
\caption{(Color online) The DOS and imaginary part of the self-energy for
  $U/W=0.6$, and $U/W=1$ for different fillings. The temperature of the system 
  is $T/W=0.05$. All data shown corresponds to the normal  phase.\label{compare_diff_doped}}    
\end{figure}

\begin{figure*}
\includegraphics[width=0.32\linewidth]{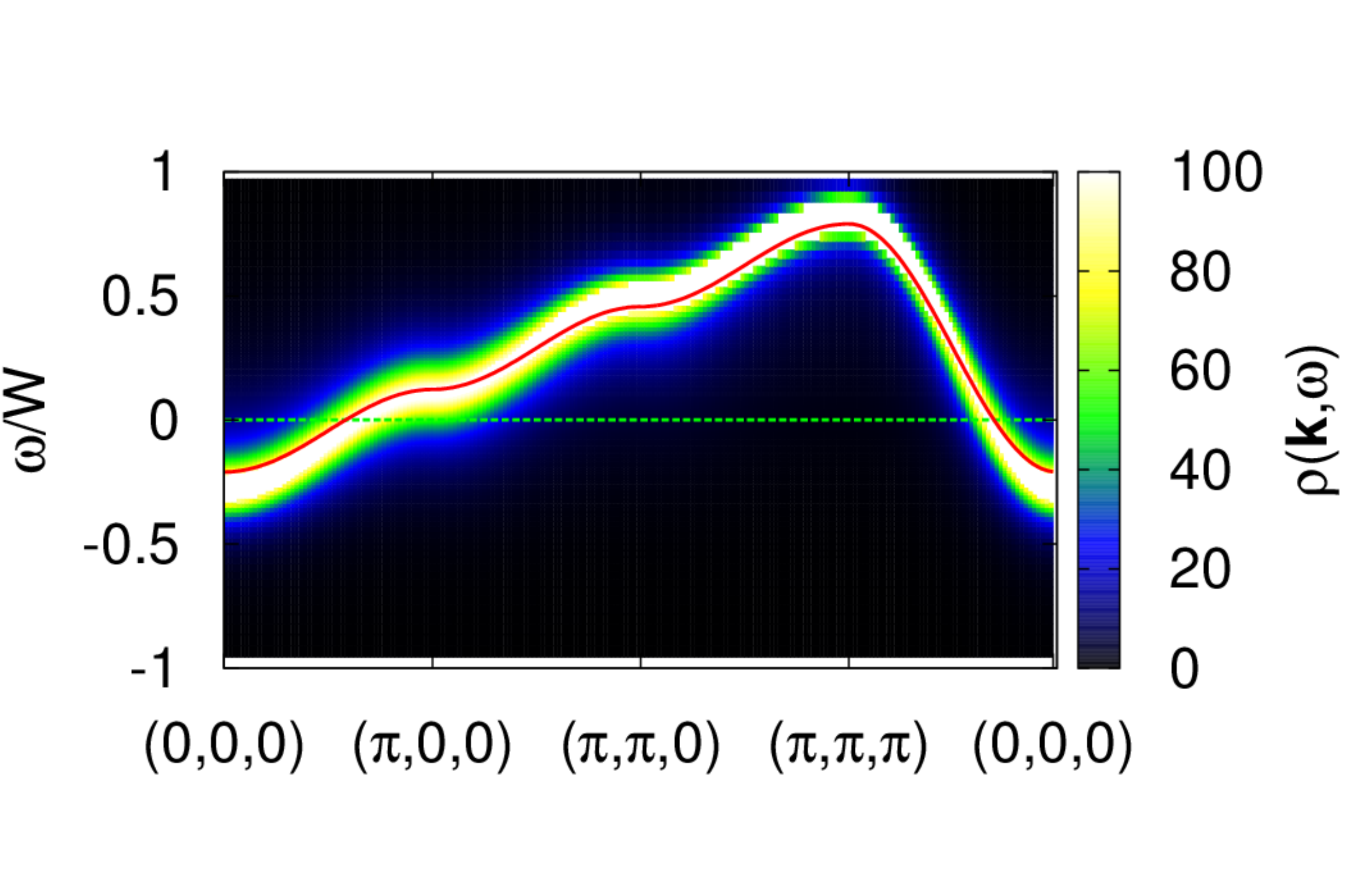}
\includegraphics[width=0.32\linewidth]{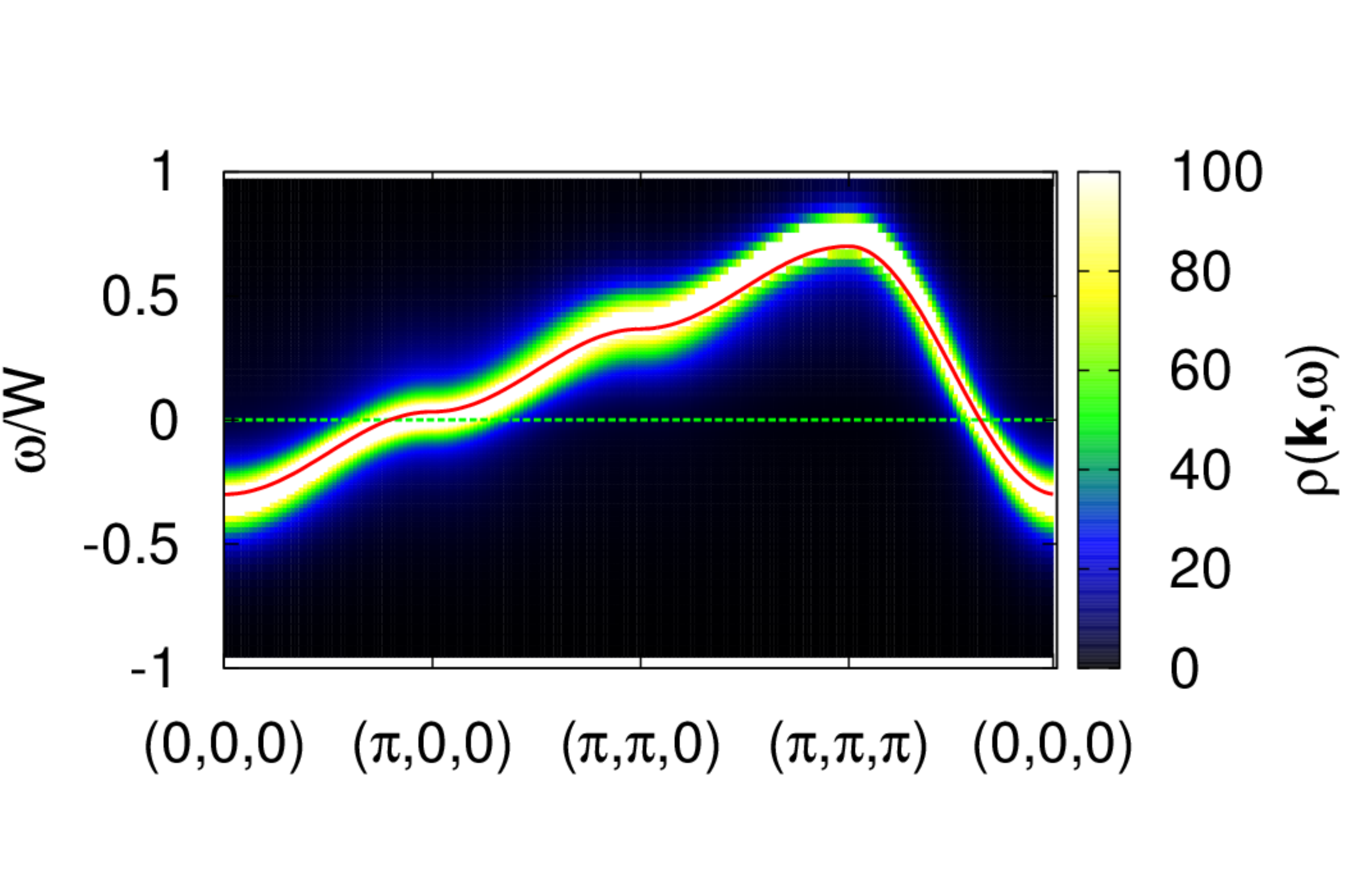}
\includegraphics[width=0.32\linewidth]{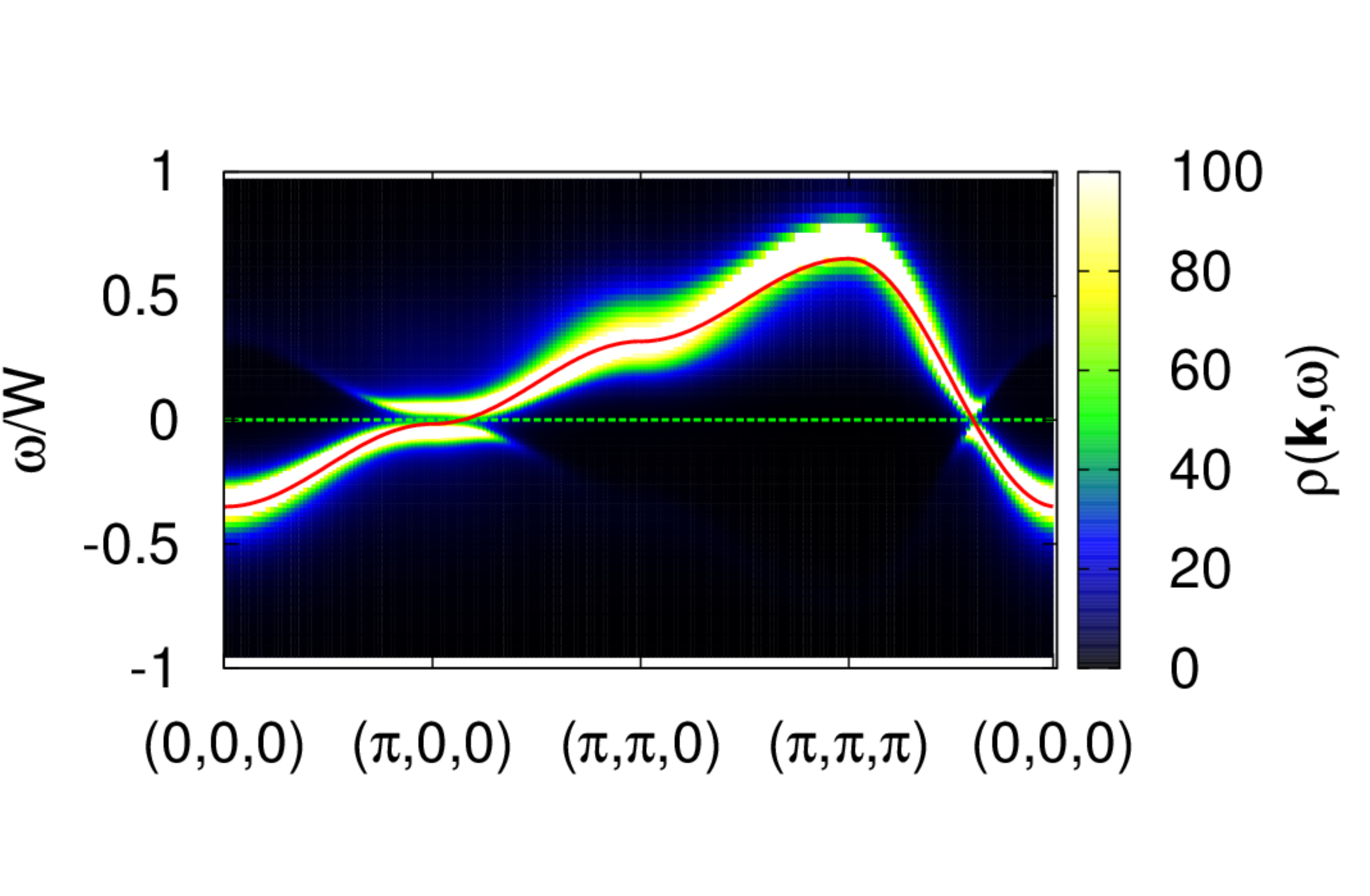}

\includegraphics[width=0.32\linewidth]{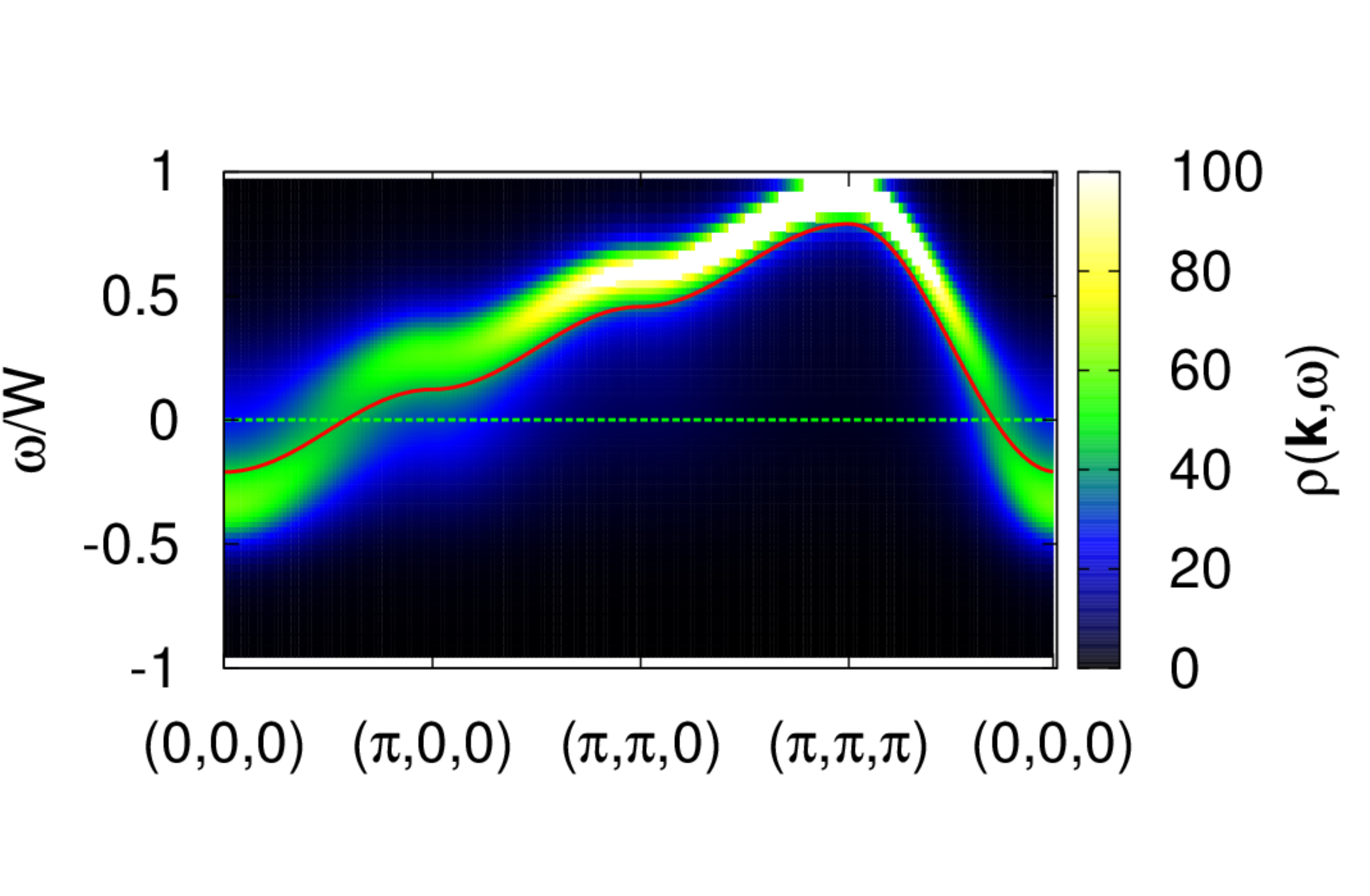}
\includegraphics[width=0.32\linewidth]{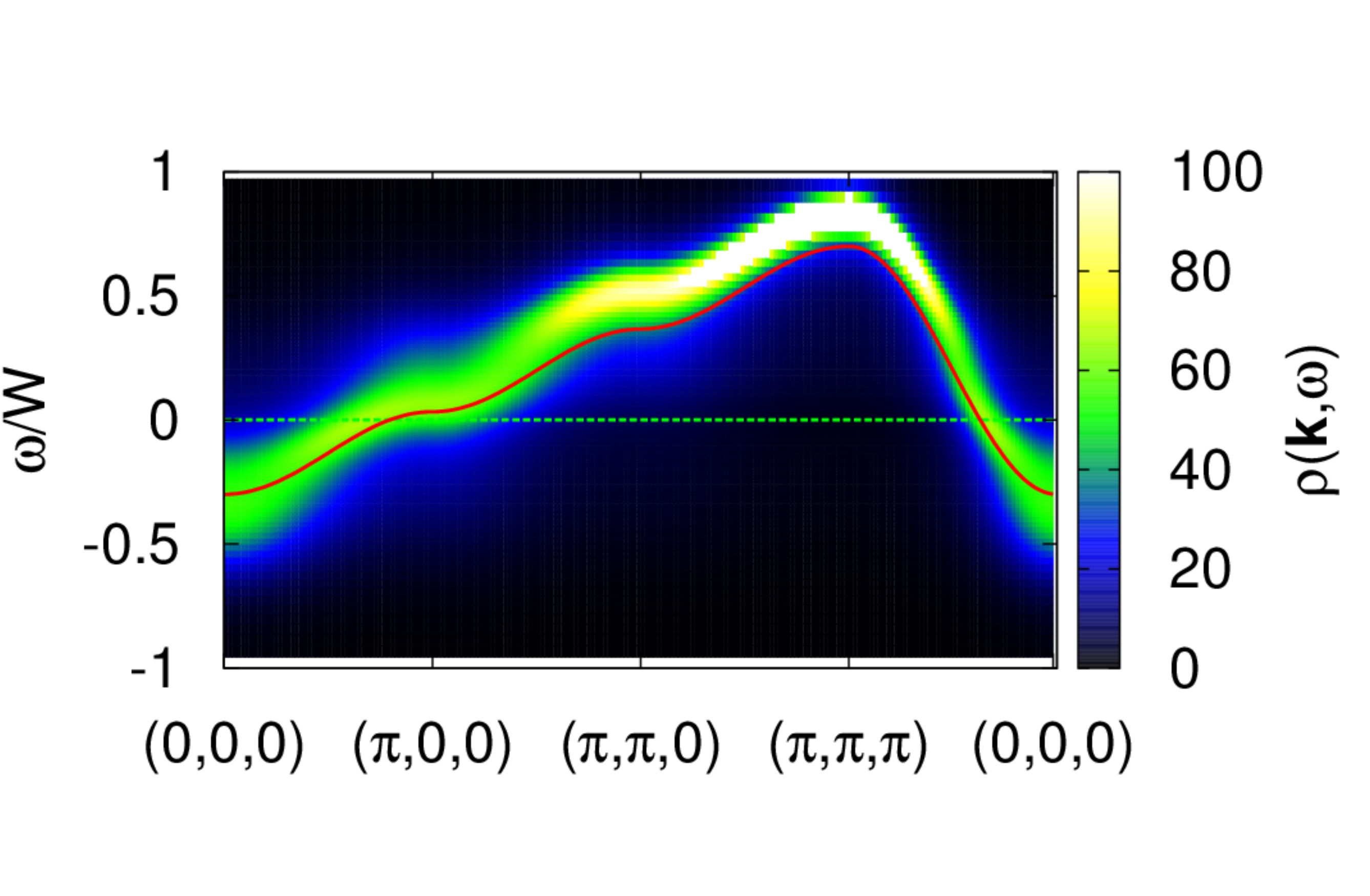}
\includegraphics[width=0.32\linewidth]{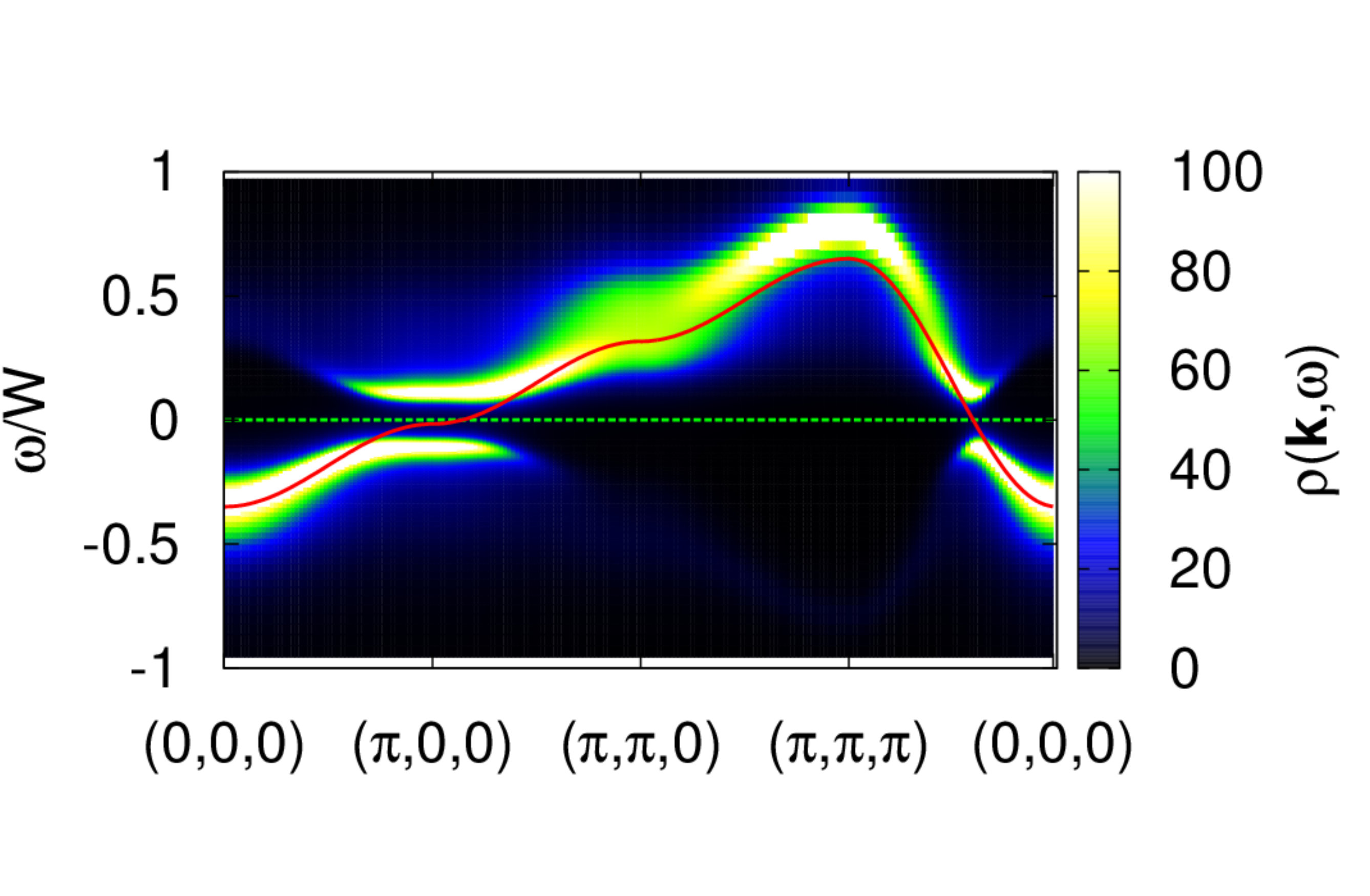}

\includegraphics[width=0.32\linewidth]{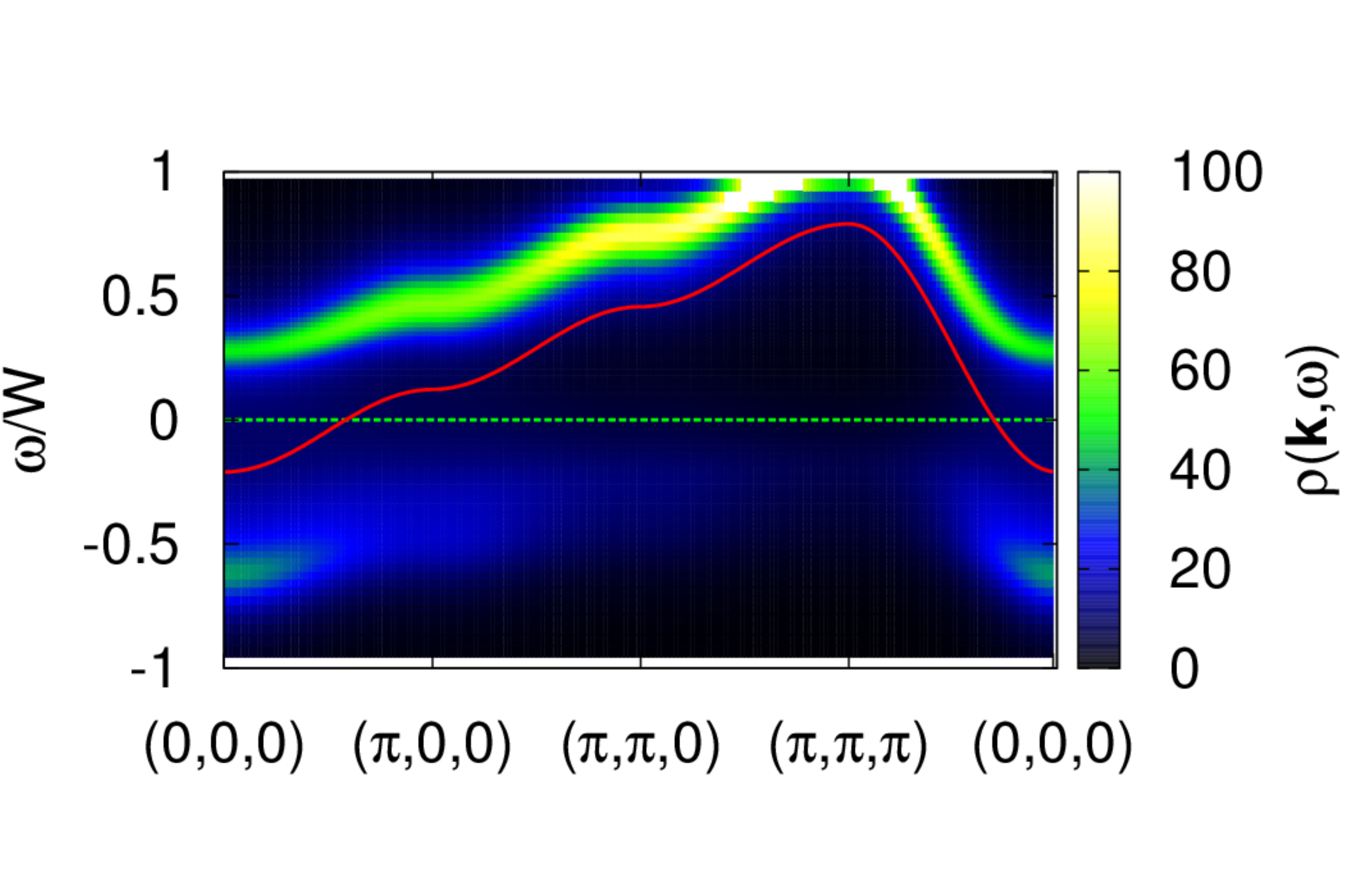}
\includegraphics[width=0.32\linewidth]{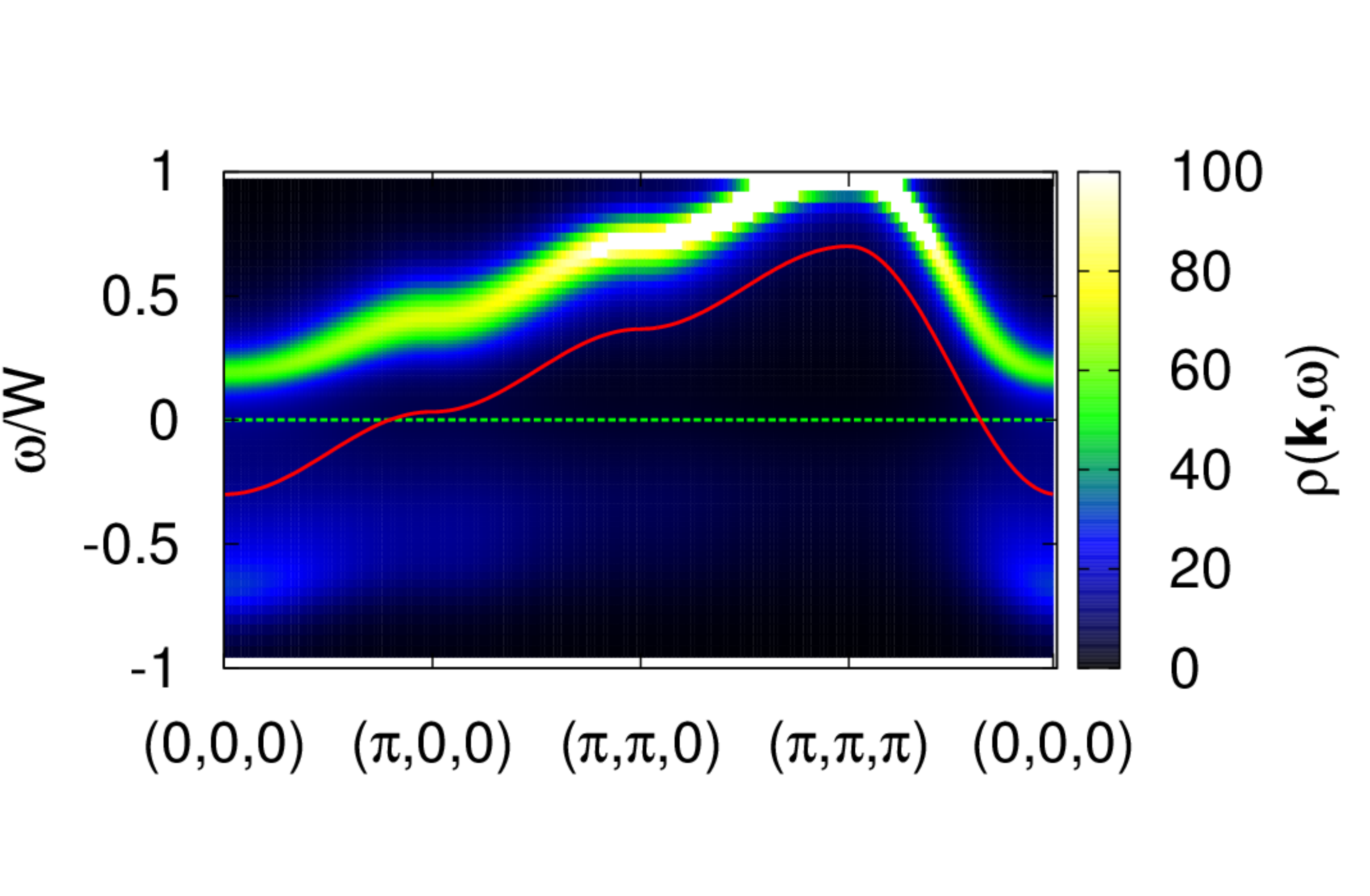}
\includegraphics[width=0.32\linewidth]{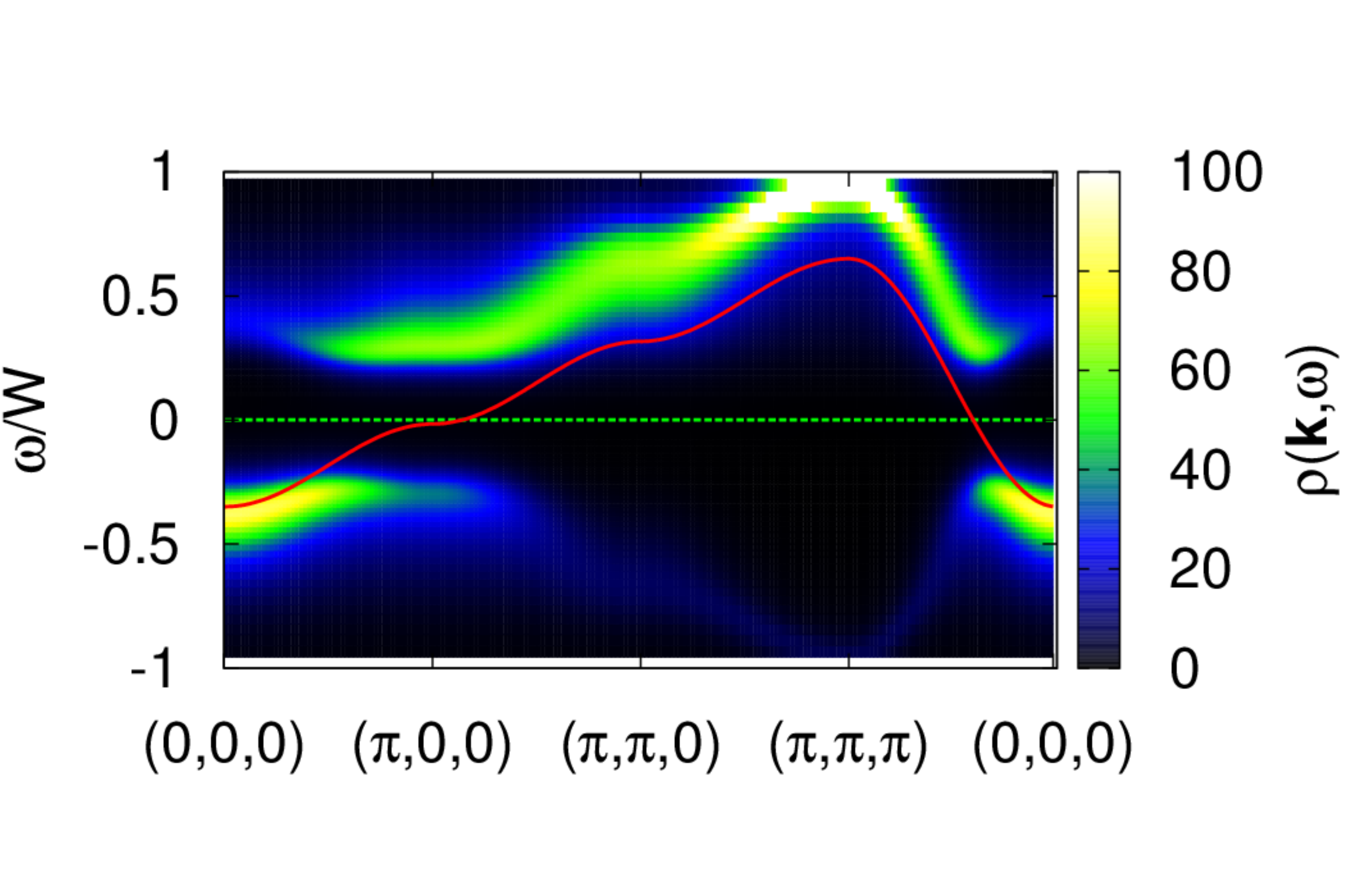}

\caption{(Color online) Momentum resolved spectral function
  $\rho_{\vk}(\omega)$ for $n=0.5$ from top to bottom  $U/W=0.4$, $U/W=0.6$, and $U/W=1$;
  and from left to right $T/W=0.2$, $T/W=0.08$, and $T/W=0.01$. The red line
  corresponds to the non-interacting dispersion $\epsilon_{\vk}$, the dashed green line
  corresponds to the Fermi energy.\label{spec_momentum_doped}}  
\end{figure*}

\section{PG physics away from half filling} 
So far we have focused on the situation at half filling where the discussion
is somewhat simplified due to the particle-hole symmetry. In this section we
show results for different filling factors ($n<1$) to see
how the PG behavior is affected. This is important for comparison with
experiments with ultracold atoms, where due to the 
trapping potential no homogeneous filling fraction can be expected.

In Fig.~\ref{compare_doped} results for $\rho(\omega)$ and
$\Imag\Sigma(\omega)$  analogous to the ones in  Fig.~\ref{compare} for the
half-filled case are displayed  for $n=0.5$ over a wide temperature range. The
lowest temperature corresponds to a gapped SF state.
Looking at the self-energies in the lower panel, we can clearly see
that the classification into FL and NFL regions is still applicable and
$|\Imag\Sigma(\omega)|$ can either show a double peak with dip in the vicinity of
$\omega=0$ (FL) or a strong single peak (NFL). It is useful here to distinguish
temperatures $T/W\lesssim 0.2$, where features are close to $\omega=0$, and
higher temperatures, where the NFL peak in $|\Imag\Sigma(\omega)|$ moves
systematically to higher energies. In contrast to the half filled situation
the case $U/W=0.6$ does not show a clear PG in $\rho(\omega)$. However, for
$U/W=1$ the PG is clearly visible. For lower temperatures the minimum in
$\rho(\omega)$ is close to $\omega=0$ and for higher temperatures it moves to
higher energies together with the NFL peak in $|\Imag\Sigma(\omega)|$. Notice,
however, that the minimum in $\rho(\omega)$ and the peak in
$|\Imag\Sigma(\omega)|$ do not coincide as they do for $n=1$.
The shift of the PG with temperature can be understood by recalling that at
high temperature the Fermi distribution becomes flatter such that higher
energies contribute to the particle number, $n=\int d\omega \rho(\omega)n_{\rm
  F}(\omega)$. To satisfy this relation at higher temperature the spectrum has
to be shifted.

The PG structure in $\rho(\omega)$ at elevated temperature can still be
understood from the local picture.  For $n<1$, we can write
$\mu=-U/2-\Delta\mu$ ($U>0$) assuming $\Delta\mu>0$,  
and the atomic energies are $E_{\alpha}=0,U/2+\Delta\mu, 2\Delta\mu$.
The partition function reads 
\begin{equation}
  Z=1+2\e^{-\beta(U/2+\Delta\mu)}+\e^{-\beta
2\Delta\mu}
\end{equation}
There are now excitations at $\omega_+=U/2+\Delta\mu$ and
$\omega_-=-U/2+\Delta\mu$ with generally asymmetric weights
\begin{equation}
 w_+=\frac{1}{Z}[1+\e^{-\beta(U/2+\Delta\mu)}],
\end{equation}
 and
 \begin{equation}
 w_-=\frac{1}{Z}[\e^{-\beta 2 \Delta\mu}+\e^{-\beta(U/2+\Delta\mu)}],
\end{equation}
respectively. 

Without showing explicit results we note that the pair density $\langle
n_\uparrow n_\downarrow\rangle$ displays a similar temperature dependence for
$n=0.5$ to what was shown in Fig.~\ref{double}, increasing from $n_{\sigma}^2$ at
large $T$ to larger values (maximal $n/2$). Therefore, similarly to the
half filled case PG behavior can coincide with an enhanced pair density for
large interactions and $T_c\lesssim T$. However, we also find cases,
e.g. $U/W=0.6$, $T/W=0.05$, with enhanced pair density ($\langle
n_\uparrow n_\downarrow\rangle \approx 0.17$)  and no PG behavior, in
contrast to the expected relation in the preformed pair scenario.

In order to get an insight to overall trends, we compare several different fillings
in Fig.~\ref{compare_diff_doped}. We show $\rho(\omega)$ and $\Imag\Sigma(\omega)$
for $U/W=0.6$ and $U/W=1$ for low temperature, $T/W=0.05$, in the normal phase.
For $U/W=0.6$ $\rho(\omega)$ exhibits a FL dip in $\Imag\Sigma(\omega)$.  It
is clearly visible, even for $n=0.1$, that the self-energy does not change its
structure when reducing the filling further. The frequency dependence in this
FL regime is relatively symmetric with respect to $\omega=0$.
The DOS, on the other hand, changes with $n$. While for $n =0.5$ a clear
peak close to the Fermi energy is visible in the DOS at low temperature, such
a peak is hardly noticeable for $n=0.2$, and it has disappeared for $n=0.1$. The
amplitude of the self-energy has become too weak to change the spectrum and we
essentially see a shifted non-interacting DOS. 

For $U/W=1$, we find a NFL peak in $\Imag\Sigma(\omega)$ for all fillings in
Fig.~\ref{compare_diff_doped}. We observe  similar effects as for weaker
interactions when reducing the filling as far as the strength of the self-energy is concerned. 
However, we find a clear PG structure in the DOS.
Whilst the PG structure at low temperature is pinned to
$\omega=0$, at high temperature the whole spectrum including the PG is shifted
to high frequencies (see Fig.~\ref{compare_doped}). 
Note that at high temperature due to the flattening of $n_{\rm F}(\omega)$  the
Fermi energy ($\omega=0$) does not play such an important role as it does
for low temperatures.

In summary, when analyzing $\rho(\omega)$ and $\Imag\Sigma(\omega)$ we can
find similar features to the ones of the half filled situation and a PG 
appears for suitable parameters. However, depending on filling, temperature,
and interaction strength, the occurrence of the PG may be limited. At high
temperatures it can be shifted away from $\omega=0$, although it is still
clearly visible in the spectrum. Moreover, the impact of the local
Hubbard interaction becomes weaker for a system with smaller filling factor.

Momentum resolved spectra for $n=0.5$ and various values of $T$ and $U$ are displayed in
Fig.~\ref{spec_momentum_doped}. Generally, the features are similar to the
half filled case. For weak coupling ($U/W=0.4$)  we find a shifted and
broadened spectrum which shows a SF gap at low temperature. 
For intermediate coupling ($U/W=0.6$) interaction effects are more visible in
the spectrum, resulting in stronger band renormalization effects and shifts of
spectral weight. However, in contrast to $n=1$ no clear PG becomes visible in
$\rho_{\vk}(\omega)$. 
For $U/W=1$, we see strong interaction effects and PG features at all
temperatures above $T_c$. We also clearly observe an asymmetry in the
intensity, which is substantially lower for the $\omega<0$ part of the spectrum.

\begin{figure}[tb]
\begin{center}
\includegraphics[width=0.95\columnwidth]{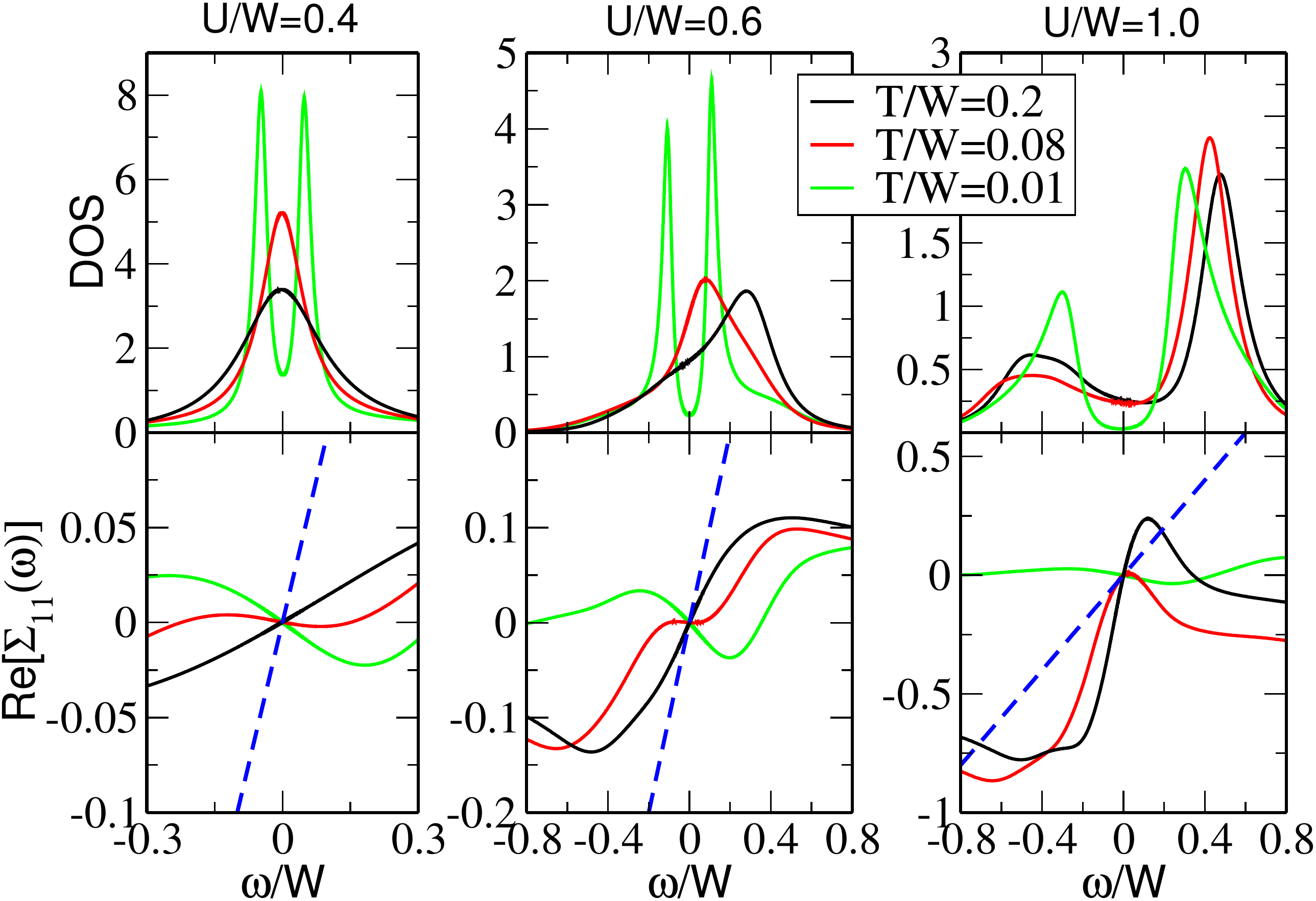}
\end{center}
\caption{(Color online) $\rho_{\vkF}(\omega)$ for $U/W=0.4$,   $U/W=0.6$ and $U/W=1$ and different
  temperatures. Notice that the lowest temperature $T/W=0.01$ is below $T_c$
  in all cases.
  The lower panels show the real part of the self-energy including the line
  $y=\omega$. \label{Gk_doped}}   
\end{figure}

Particular cuts along $\omega$ for momenta which satisfy $\xi_{\vkF}+\Real\Sigma(0)=0$ are
shown in Fig.~\ref{Gk_doped}. 
At weak coupling ($U/W=0.4$) the FL peak is gapped out when the temperature is
lowered below $T_c$. For intermediate coupling ($U/W=0.6$) above $T_c$ we find
that the FL peak is shifted away from $\omega=0$ to higher energies. Also the
coherence peaks below $T_c$ show some asymmetry due to self-energy effects.
A clear PG is only visible for larger interactions, $U/W=1$. The lower panel
shows again the real part of the self-energy. In 
contrast to the situation at half filling, the peaks in $\rho_{\vkF}(\omega)$
are not well explained by the intersection, $\omega=\Real\overline{\Sigma}(\omega)$. In
this situation the variation of $\Imag\Sigma(\omega)$ is too strong invalidating
the simple arguments of Sec.~III. Nevertheless a NFL peak form of the
self-energy is clearly important for the PG behavior.

\section{Discussion and Conclusions}

We have analyzed the occurrence of PG features in the integrated and
$\vk$-resolved spectral function of the three-dimensional attractive Hubbard
model for different temperatures, interactions and filling factors. Properties
of the spectral functions have been traced back to the characteristic behavior
of the self-energy. We find  PG behavior as long as the interaction $U$ is
large enough ($\sim W$) and the self-energy shows NFL behavior, i.e., $T>
T_{\rm FL}(U)$.  
Our results show marked deviations from the popular preformed pair scenario,
where PG behavior is directly linked to the formation of pairs at a temperature
$T_{\rm p}>T_c$: (i) We find that PG behavior persists up to large temperatures
and is not bounded by some temperature scale $T_{\rm p}$. (ii) We find cases
with a substantially enhanced pair density where no PG behavior occurs. 
The first effect is related to the fact that we are working with a lattice
model, such that local excitations are always well defined and related to
the chemical potential $\mu$ and $U$. This might be different in the continuum
where it is conceivable that the preformed pair scenario of Fig.~1 is
applicable.  On the other hand we expect the PG to be present at large
temperatures as a non-perturbative local lattice effect also  in
the two-dimensional lattice model. 
Certainly, other effects like strong phase fluctuations and small momentum
pairing fluctuations, not contained in our calculations, can lead to an
extension of the regimes where PG behavior occurs.

A word of caution is in order when discussing the large temperatures addressed
in this paper. Here we dealt with a strict one-band model where the kinetic energy is limited by
the bandwidth. In most real systems very high temperature would activate
higher bands, and in solid state systems, it can lead to the melting of the
crystal structure; such effects are obviously not allowed in our setup.

Experiments with ultracold atoms in optical lattices provide an excellent
platform to test our predictions. Interactions can be tuned in a wide range by
Feshbach resonances, the lattices can be loaded with different filling factors
and a temperature range $T/W=0.1-0.2$ is routinely accessible
\cite{BDZ08}. Integrated and momentum resolved spectra can be measured such
that a direct comparison with our predictions is possible.
Thus, we hope that our work will stimulate further efforts in this field which contribute to a
better understanding of the intriguing PG physics.

\paragraph*{Acknowledgments -} We wish to thank M. Capone, N. Dupuis, 
A. Georges, O. Gunnarsson, B. Halperin, A. Koga, W. Metzner, E. Perepelitsky, M. Punk,
P. Strack, and A. Toschi for very helpful discussions and suggestions during
different stages of this work. JB acknowledges financial support from the DFG  
through grant number BA 4371/1-1.  RP is supported by the FPR program of
RIKEN. Computer calculations have been done at the RICC supercomputer at RIKEN
and the Kashiwa supercomputer of the Institute of Solid State Physics in
Japan.

\begin{appendix}
\section*{Appendix}

\subsection{T-matrix approximation}
A popular approximation for the self-energy is the so-called $T$-matrix
approximation, which corresponds essentially to summing the scattering
processes in the particle-particle channel. One has,\cite{Hau92,KMS99}
\begin{equation}
  \Sigma^{(1)}=TU\sum_{m,\vq}\e^{i\omega_m\eta}
  G(\vq,i\omega_m) ,
\label{eq:sigmaiw}
\end{equation}
or equivalently,
\begin{equation}
  \Sigma^{(1)}=U\sum_{\vq}\integral{\omega}{}{}\rho(\vq,\omega)n_{\rm
    F}(\omega) ,
\label{eq:sigmaiw}
\end{equation}
and
\begin{equation}
  \Sigma_{\vk}^{\rm T}(i\omega_n)=T\sum_{m,\vq}\e^{i\omega_n\eta}\Gamma(\vq,i\omega_m)
  G(\vq-\vk,i\omega_m-i\omega_n) ,
\label{eq:sigmaiw}
\end{equation}
with $\eta\to0$. Here we defined
\begin{equation}
  \Gamma(\vq,i\omega_m)=\frac{U^2K(\vq,i\omega_m)}{1-U K(\vq,i\omega_m)},
\end{equation}
with the particle-particle propagator
\begin{equation}
  K(\vq,i\omega_m)=-T\sum_{n,\vq} G(\vq-\vk,i\omega_m-i\omega_n)G(\vk,i\omega_n).
\end{equation}
The self-energy is
$\Sigma_{\vk}(i\omega_n)=\Sigma_{\vk}^{(1)}(i\omega_n)+\Sigma_{\vk}^{\rm T}(i\omega_n)$.

In the local approximation, the expression simplify. We find the following
result after analytic continuation,
\begin{equation}
  \Sigma^{(1)} =U\integral{\omega}{}{}\rho_G(\omega)n_{\rm F}(\omega) ,
\label{eq:sigmalociw}
\end{equation}
and
\begin{equation}
  \Sigma^{\rm T}(\omega)=\integral{\omega_1}{}{}\integral{\omega_2}{}{}\frac{\rho_{\Gamma}(\omega_1)
    \rho_G(\omega_2)}{\omega^+-\omega_1+\omega_2}[n_{\rm  B}(\omega_1)+n_{\rm F}(\omega_2)] .
\end{equation}

We have
\begin{equation}
  K(i\omega_m)=-T\sum_{n} G(i\omega_m-i\omega_n)G(i\omega_n),
\end{equation}
and $\rho_{\Gamma}=-\frac{1}{\pi}\Imag\Gamma(\omega^+)$.
Introducing spectral functions we can also write,
\begin{equation}
  K(\omega^+)=\integral{\omega_1}{}{}\integral{\omega_2}{}{}\frac{\rho_G(\omega_1)\rho_G(\omega_2)}
{\omega^+-\omega_1-\omega_2}[n_{\rm  F}(\omega_1)-n_{\rm F}(-\omega_2)],
\end{equation}
and
\begin{equation}
  \Gamma(i\omega_m)=\frac{U^2K(i\omega_m)}{1-U K(i\omega_m)}.
\end{equation}
The $T$-matrix calculations can be done non-self-consistently (Tnsc) and
self-consistently (Tsc).

\subsection{Comparison of NRG-DMFT with IPT and T-matrix}

\begin{figure}[!ht]
\begin{center}
\vspace{0.2cm}
\includegraphics[width=0.48\columnwidth]{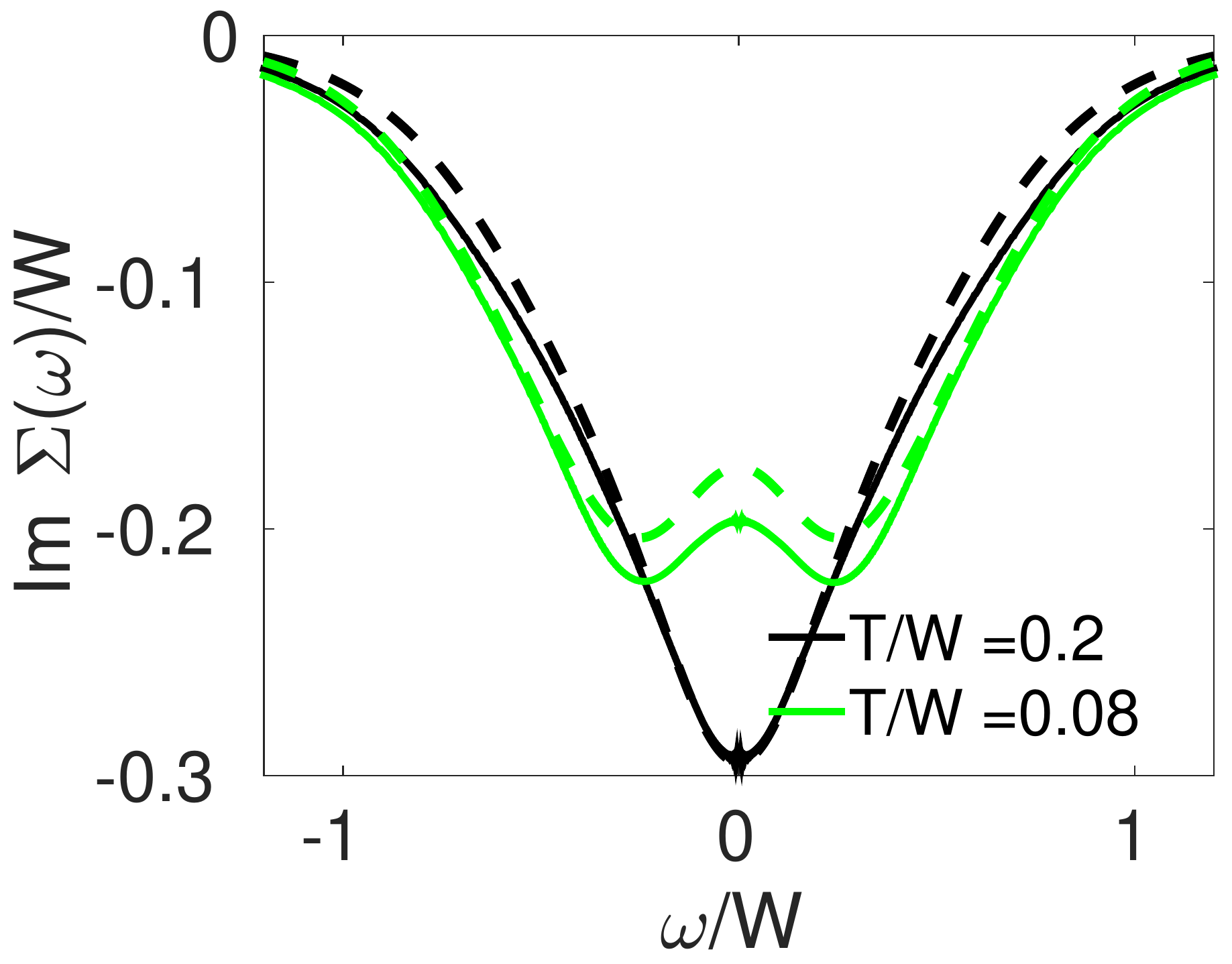}
\includegraphics[width=0.48\columnwidth]{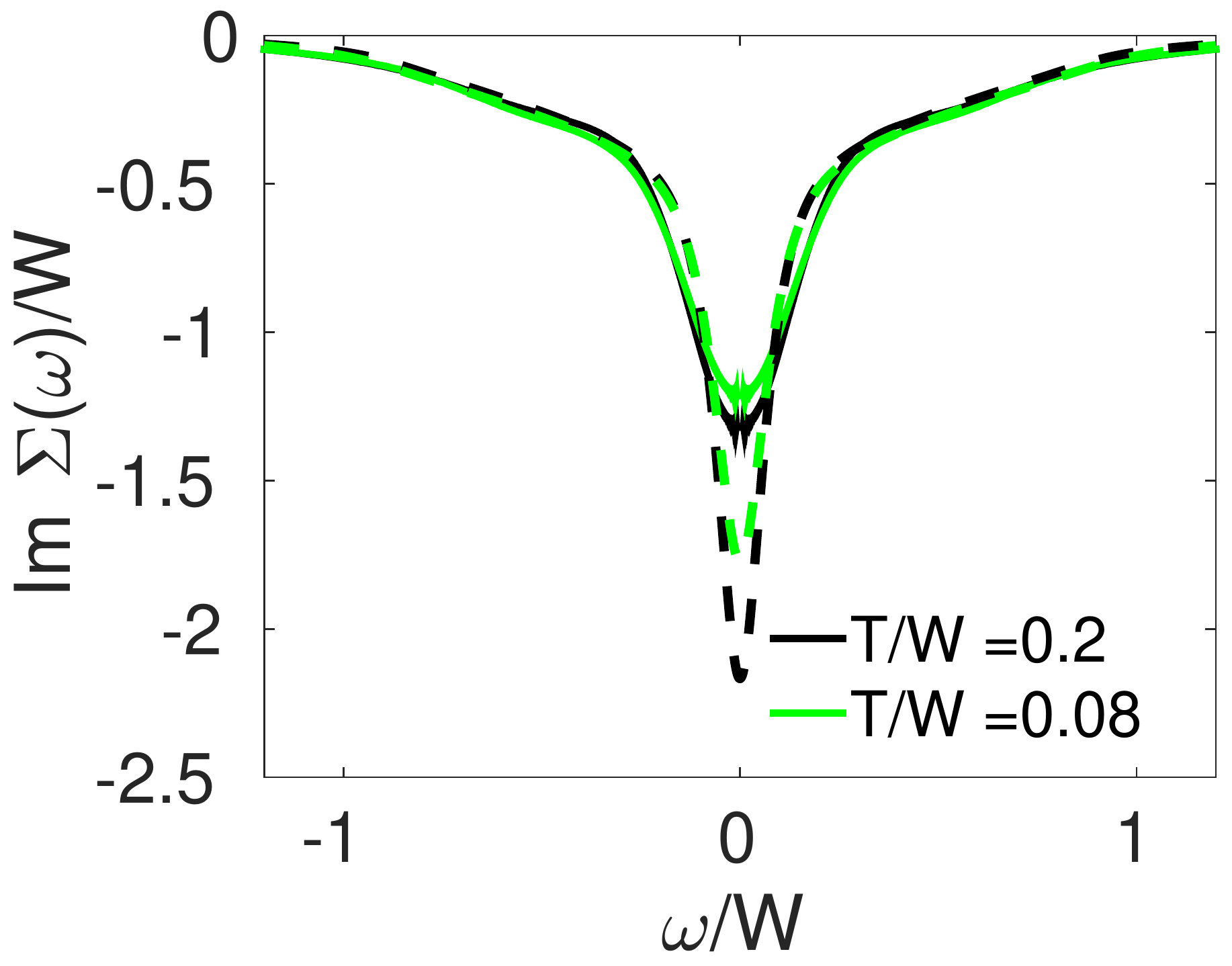}
\includegraphics[width=0.48\columnwidth]{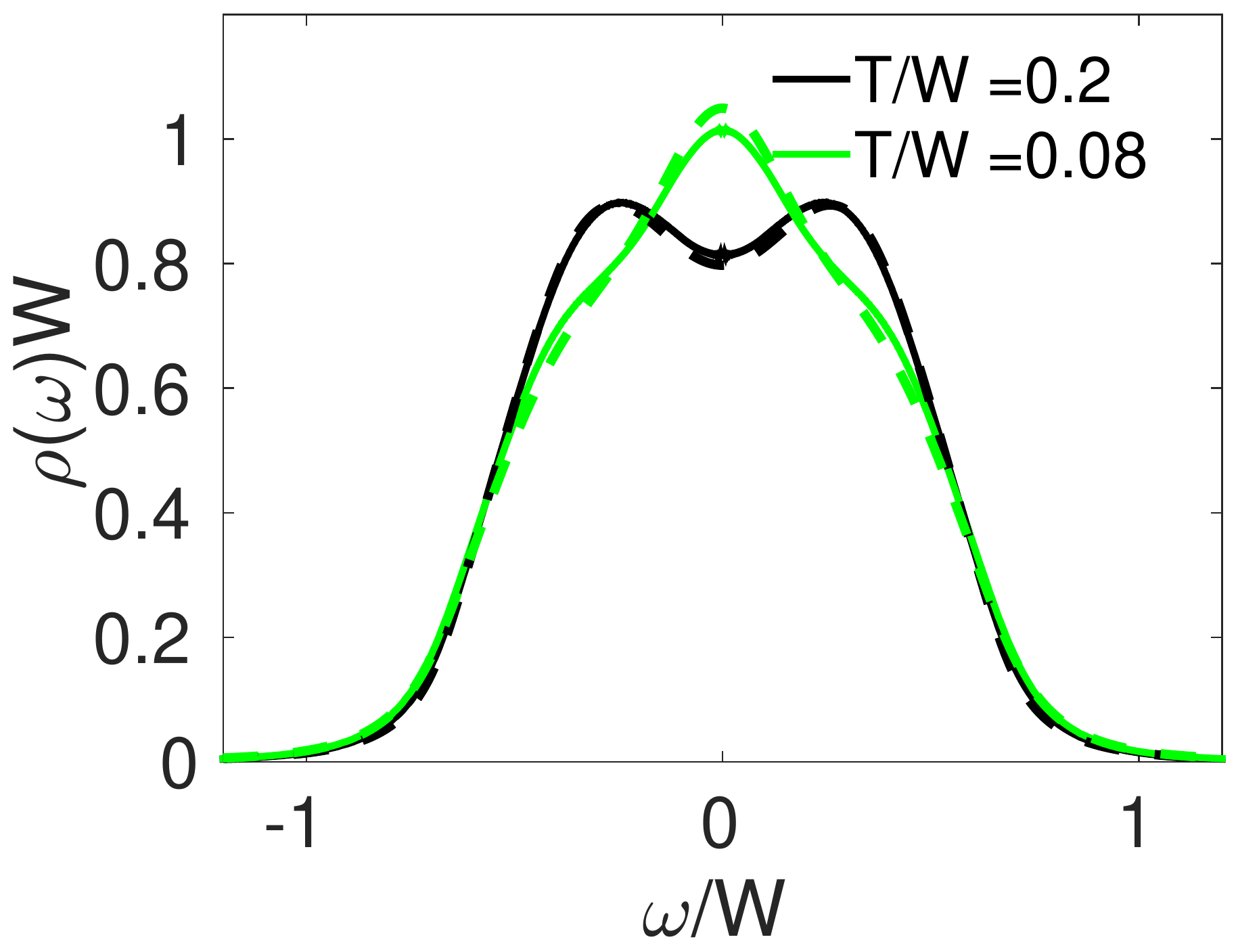}
\includegraphics[width=0.48\columnwidth]{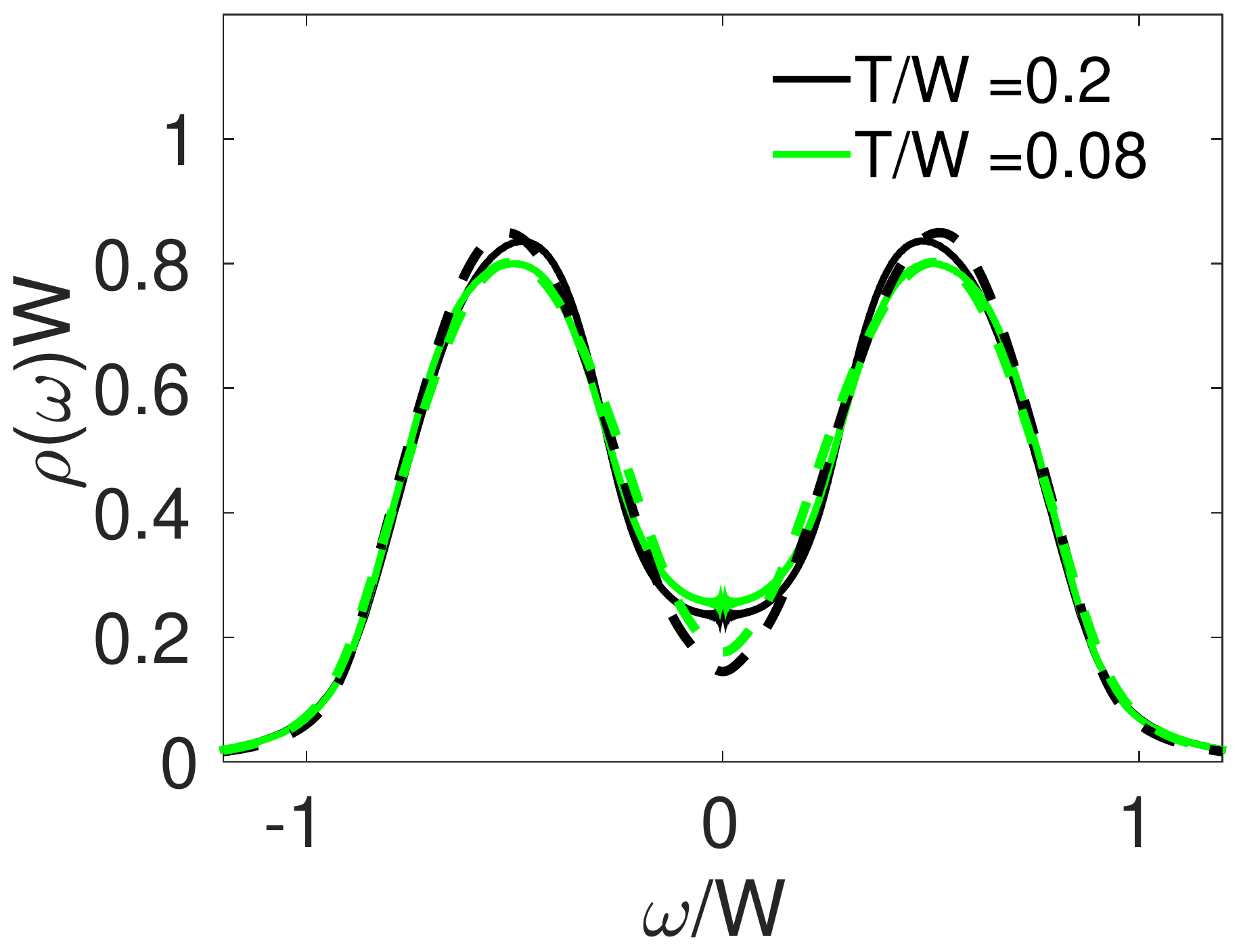}
\end{center}
\vspace{-0.5cm}
\caption{(Color online) Comparison of DMFT-NRG (full lines)
  and IPT (dashed lines)  results for $U/W=0.6$ (left) and $U/W=1$ (right) for
  $\Imag\Sigma (\omega)$ and $\rho(\omega)$.
\label{comp_ipt}}
\end{figure}

We start with a comparison of the DMFT results obtained using NRG calculations
for the effective impurity model with DMFT calculations using second order
perturbation theory, usually termed iterated perturbation theory (IPT). IPT gives
qualitatively reliable results in the half filled Hubbard model.\cite{GKKR96}
Since IPT does not require a prescription of broadening discrete excitations,
this comparison helps to validate the finite temperature broadening procedure 
described in Sec.~II. We focus on results at half filling in this section.
In Fig.~\ref{comp_ipt} we show a comparison of the imaginary part of the
self-energy, $\Imag\Sigma (\omega)$, and the integrated spectral function,
$\rho(\omega)$, for $U/W=0.6$ (left) and $U/W=1$ (right) and different temperatures.

\noindent
Overall the agreement is good with minor deviations in the tails. There is
a particularly visible difference for $U/W=1$, where the IPT result for
$\Imag\Sigma (\omega)$ shows a somewhat stronger peak. This leads to a more
pronounced PG in $\rho(\omega)$. 
We conclude that the DMFT-NRG results at high temperatures have the
qualitative correct form and the PG remains there. 

We also provide a comparison of the DMFT-NRG results with T-matrix
calculations. In particular, we use the Eq.~(\ref{eq:sigmalociw}) and
following, and include self-consistent (Tsc) and non-selfconsistent 
(Tnsc) results. Note that the T-matrix calculations are only sensible as long as 
$1-U \Real K(\omega)$ does not become zero, which is particularly important for
the non-selfconsistent case.
At weak coupling ($U/W= 0.2$, not shown) one can find reasonable agreement of
T-matrix calculations with the DMFT-NRG and all calculations give no PG
behavior. However, in this situation also second order perturbation theory
gives satisfactory agreement. 

\begin{figure}[!ht]
\begin{center}
\vspace{0.2cm}
\includegraphics[width=0.48\columnwidth]{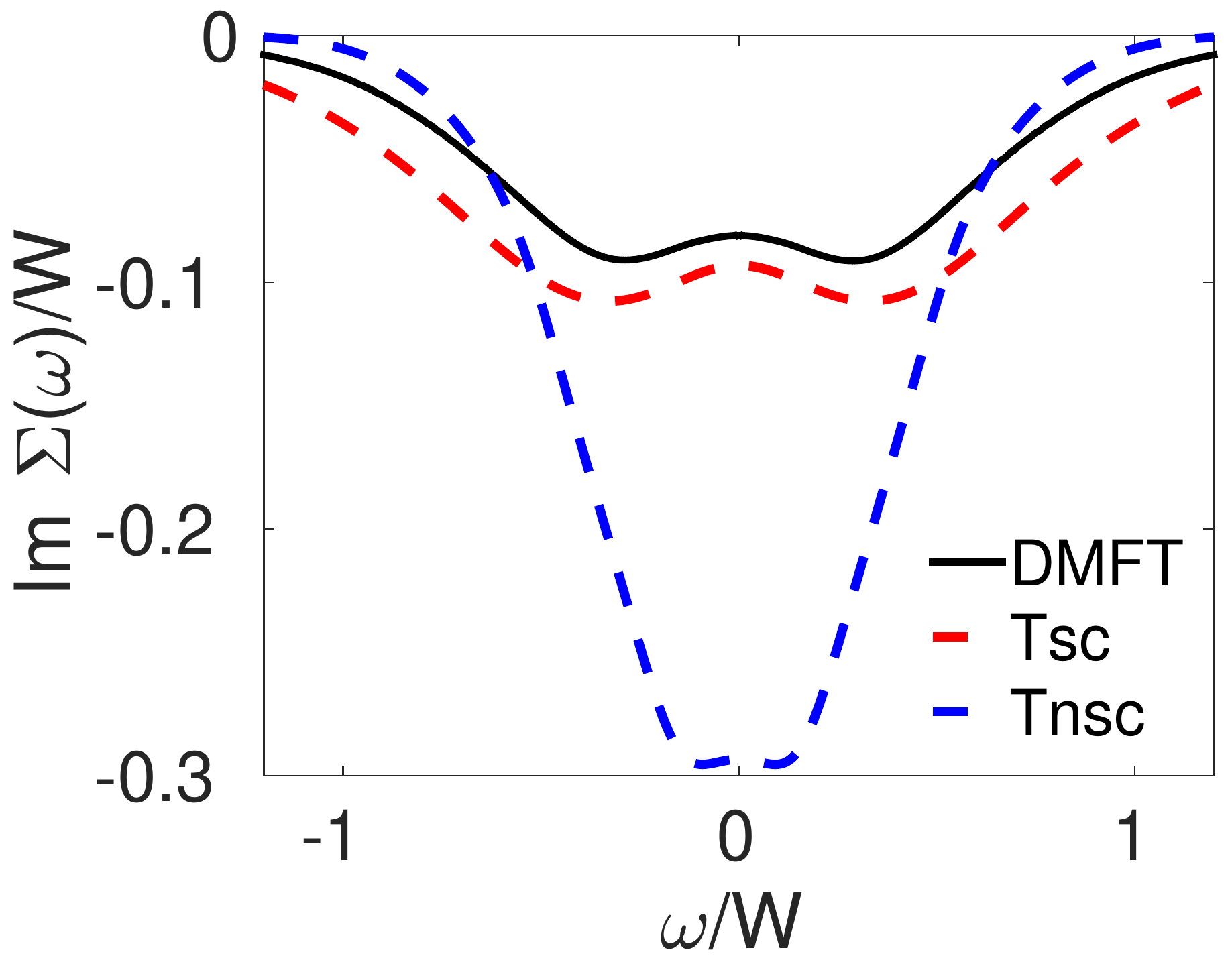}
\includegraphics[width=0.48\columnwidth]{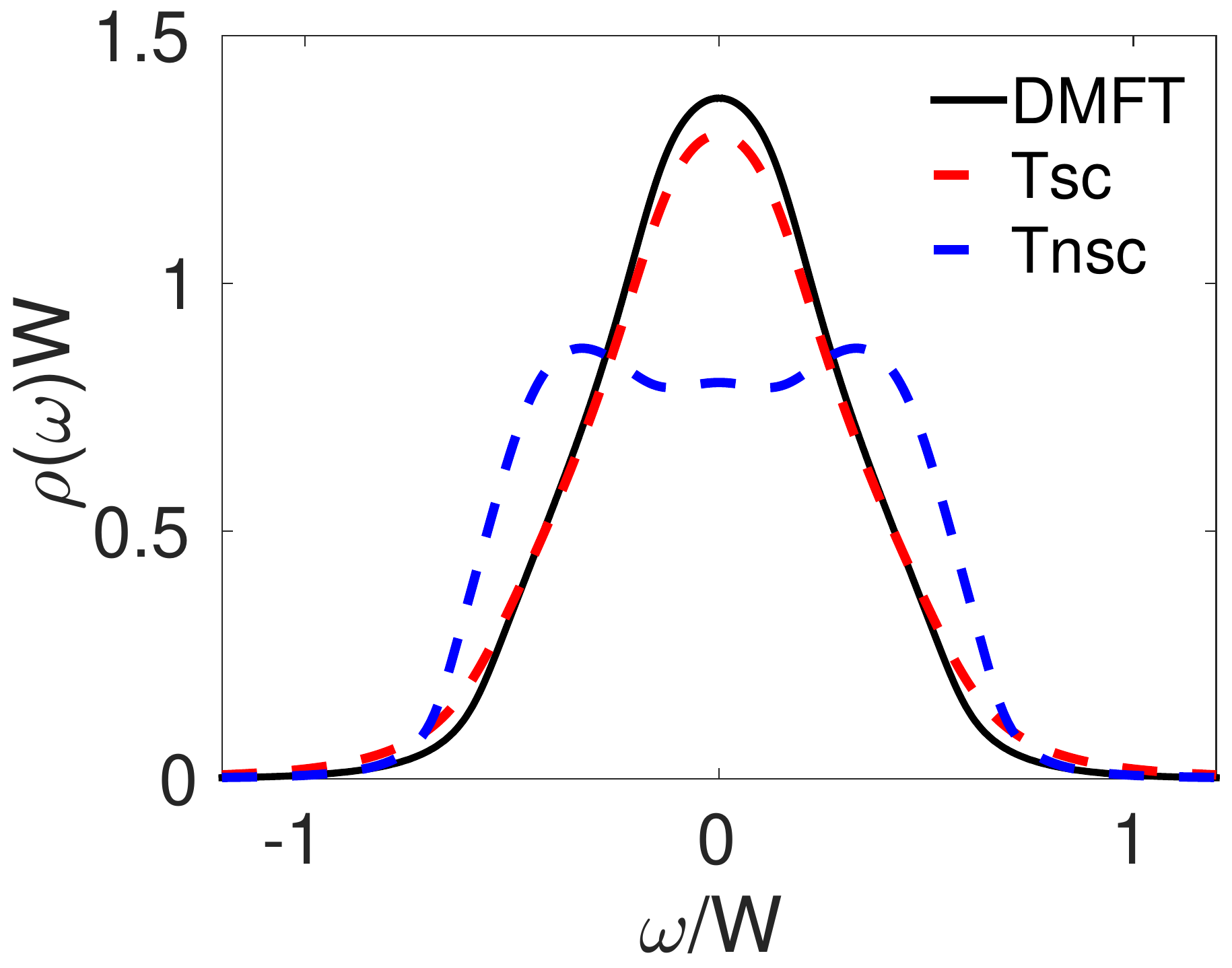}
\end{center}
\vspace{-0.5cm}
\caption{(Color online) Comparison of DMFT-NRG result with self-consistent
  (Tsc) and non-selfconsistent (Tnsc) T-matrix calculations for $\Imag\Sigma
  (\omega)$ and $\rho(\omega)$ for $U/W=0.4$ and $T/W=0.1$.}
\label{comp_tmatU04}
\end{figure}
\noindent
As seen in Fig.~\ref{comp_tmatU04} for $U/W= 0.4$ and $T/W=0.1$, Tsc and DMFT still show
reasonable agreement, whereas Tnsc calculations can lead to a strong
overestimate for $\Imag\Sigma (\omega)$. This can lead to a PG feature in $\rho(\omega)$, even though
calculations with the DMFT-NRG give no PG behavior.

\begin{figure}[!ht]
\begin{center}
\vspace{0.2cm}
\includegraphics[width=0.48\columnwidth]{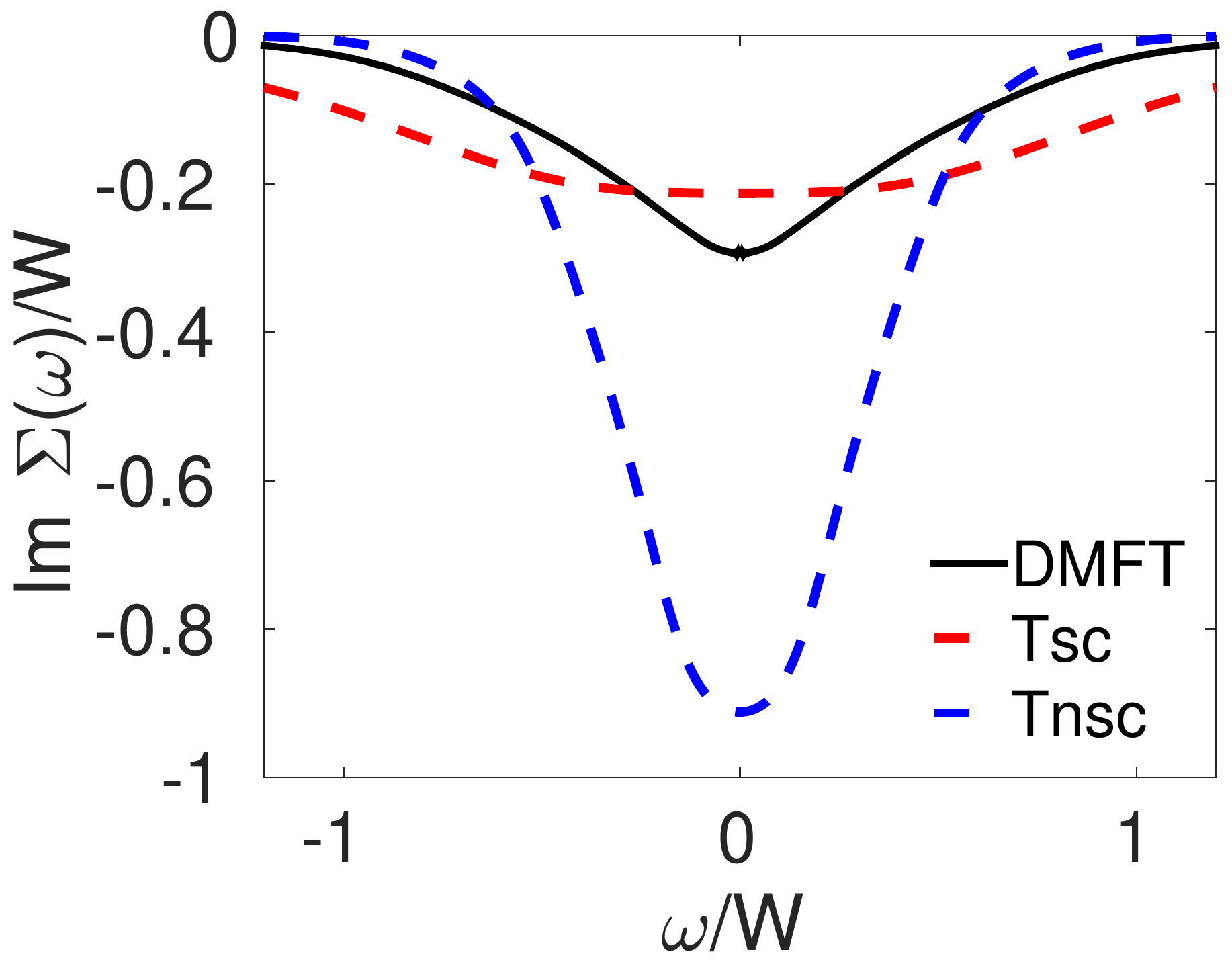}
\includegraphics[width=0.48\columnwidth]{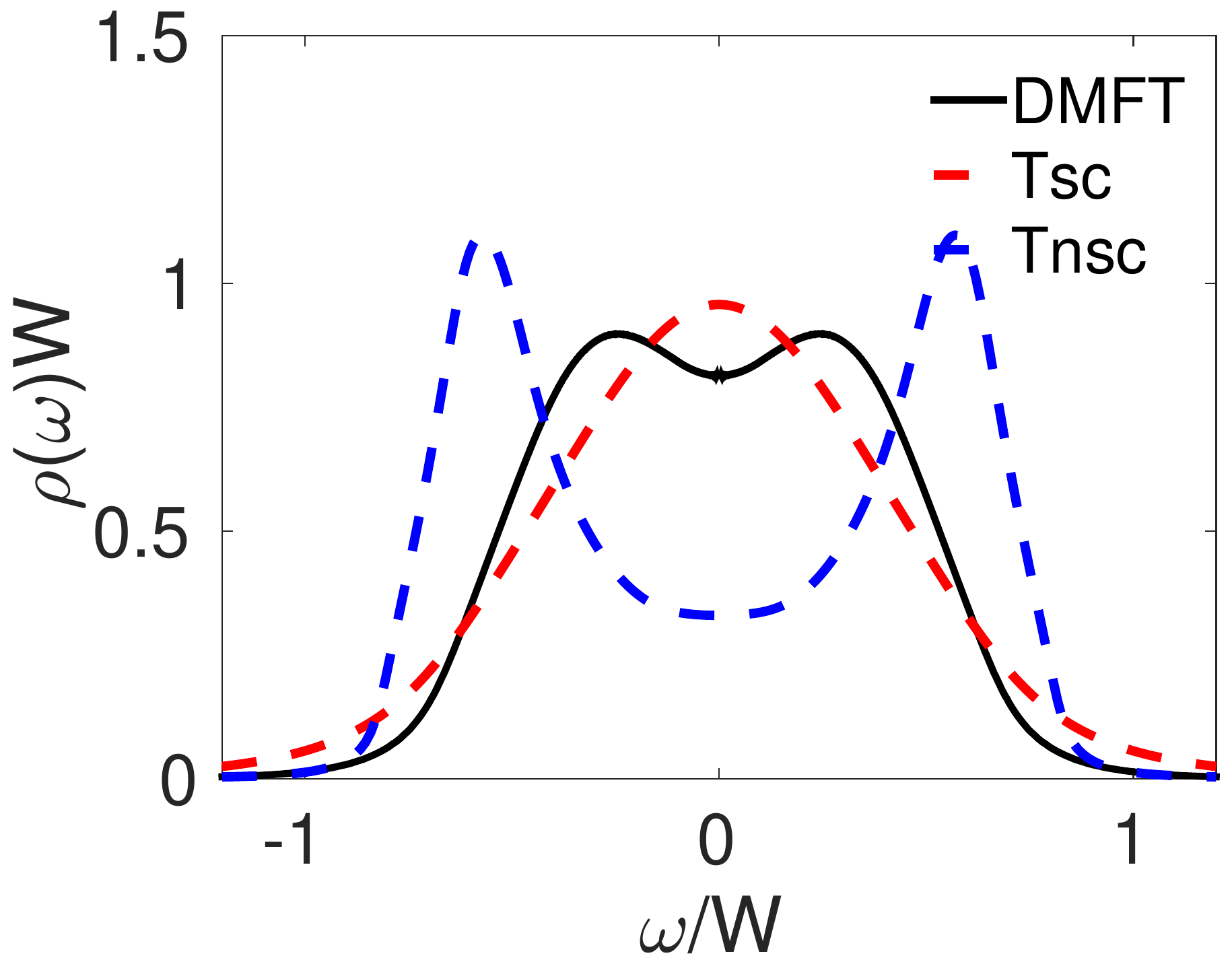}
\end{center}
\vspace{-0.5cm}
\caption{(Color online) Comparison of DMFT-NRG result with self-consistent
  (Tsc) and non-selfconsistent (Tnsc) T-matrix calculations for $\Imag\Sigma
  (\omega)$ and $\rho(\omega)$ for $U/W=0.6$ and $T/W=0.2$.
\label{comp_tmatU06}}
\end{figure}

For intermediate coupling, $U/W=0.6$, and $T/W=0.2$, we show a further comparison
in Fig.~\ref{comp_tmatU06}. 

\noindent
In this case both T-matrix calculations give unreliable results. The
self-energy of the self-consistent version is too small and $\rho(\omega)$
shows no PG. The non-selfconsistent calculation shows a PG but its magnitude
is largely overestimated.
For larger interactions, for instance, $U/W=1$,  the deviations get worse. We
therefore conclude that T-matrix calculations - both self-consistent and
non-selfconsistent - within the local approximation do not give reliable
results for the PG physics of the three dimensional Hubbard model at half
filling.

\subsection{Second order self-energy and phase space factor}

The result for the second order retarded self-energy reads,\cite{SC91}
\begin{equation}
  \Sigma^r(\omega,\vk) = U^2\integral{\epsilon}{}{} 
   \frac{F^r(\epsilon,\vk)}{\omega+i\eta-\epsilon}.
\end{equation}
The imaginary part of the retarded self-energy is then given by  
\begin{equation}
\Imag\Sigma^r_{\vk}(\omega)=-\pi U^2 F^r(\omega,\vk), 
\end{equation}
where $F^r(\epsilon,\vk)=f_1(\epsilon,\vk)+f_2(\epsilon,\vk)$,
with the phase space factors, 
\begin{widetext}
\begin{equation}
f_1(\epsilon,\vk)=\sum_{\vk_1,\vk_2,\vk_3}\delta(\xi_{\vk_2}+\xi_{\vk_3}-\xi_{\vk_1}-\epsilon)
\delta(\vk+\vk_1,\vk_2+\vk_3)n_{\vk_1}(1-n_{\vk_2})(1-n_{\vk_3}).
\end{equation}
and,
\begin{equation}
f_2(\epsilon,\vk)=\sum_{\vk_1,\vk_2,\vk_3}\delta(\xi_{\vk_2}+\xi_{\vk_3}-\xi_{\vk_1}-\epsilon)
\delta(\vk+\vk_1,\vk_2+\vk_3)(1-n_{\vk_1})n_{\vk_2}n_{\vk_3}.
\end{equation}
The expressions can be simplified in the limit of large dimensions. The momentum
integrations can be replaced by integrals over the density of states, momentum
conservation is implicit so we can omit the corresponding $\delta$-function
and the $\vk$-dependence disappears,
\begin{equation}
f_1(\epsilon)=\integral{\epsilon_1}{}{}\!\integral{\epsilon_2}{}{}\!\integral{\epsilon_3}{}{}
\rho_0(\epsilon_1)\rho_0(\epsilon_2)\rho_0(\epsilon_3) 
\delta(\epsilon_2+\epsilon_3-\epsilon_1-\epsilon-\mu)n_{\rm
  F}(\epsilon_1-\mu)n_{\rm F}(-\epsilon_2+\mu)n_{\rm F}(-\epsilon_3+\mu). 
\end{equation}
We can do the integration over the $\delta$-function,
\begin{equation}
f_1(\epsilon)=\integral{\epsilon_2}{}{}\!\integral{\epsilon_3}{}{}
\rho_0(\epsilon_2+\epsilon_3-\epsilon-\mu)\rho_0(\epsilon_2)\rho_0(\epsilon_3) 
n_{\rm  F}(\epsilon_2+\epsilon_3-\epsilon-\mu)n_{\rm F}(-\epsilon_2+\mu)n_{\rm F}(-\epsilon_3+\mu),
\end{equation}
\end{widetext}
and similarly for $f_2(\epsilon)$.
In the particle-hole symmetric case we have,
\begin{equation}
f_2(\epsilon)=f_1(-\epsilon),
\end{equation}
It is then sufficient to evaluate $f_1(\epsilon)$ and we can
write, 
\begin{equation}
  F^r(\epsilon,\vk)=F^r(\epsilon)=f_1(\epsilon)+f_1(-\epsilon).
\end{equation}
This can be evaluated as a double integral for a given temperature and
$\rho_0(\epsilon)$.
Assuming that $\rho_0(\epsilon)$ is only finite in an interval $(-D,D)$ we can
analyze the double integration as being determined by certain region in the
$\epsilon_3-\epsilon_2$ plane.
At $T=0$ a geometric analysis of the integration region shows,
$f_1(\epsilon)\sim \epsilon^2$, which gives the typical Fermi liquid behavior, Eq.~(\ref{eq:FL}),
at low temperature. 
In the opposite limit, $T\to \infty$, a similar analysis shows that
$F^r(\epsilon)$ is maximal at $\epsilon=0$ and it decays for small $\epsilon$
as $-\epsilon^2$, which yields the NFL form Eq.~(\ref{eq:NFL}). One can estimate the
crossover temperature $T_{\rm FL}$ by studying when then coefficient of the
$\epsilon^2$ changes sign. Depending on the density of states and the
approximations made one finds a result of the order of a fraction of the
bandwidth, consistent with the result in Fig.~\ref{phase_diagram} for small
$U$.

\end{appendix}

\bibliography{artikel,biblio1,footnote}

\begin{thebibliography}{71}
\bibitem[{fn1()}]{fn1}
\bibinfo{note}{We would like to thank N. Dupuis for pointing this out to us.}
\bibitem[{fn2()}]{fn2}
\bibinfo{note}{We use this as a general condition also at finite temperature,
  even though the Fermi surface is strictly only defined at $T=0$.}
\end{thebibliography}

\end{document}